\documentclass[nonacm,manuscript,screen]{acmart}

\AtBeginDocument{%
  }

\setcopyright{acmlicensed}
\copyrightyear{2018}
\acmYear{2018}
\acmDOI{XXXXXXX.XXXXXXX}

\acmConference[Conference acronym 'XX]{Make sure to enter the correct
  conference title from your rights confirmation emai}{June 03--05,
  2018}{Woodstock, NY}
\acmISBN{978-1-4503-XXXX-X/18/06}
\usepackage{xspace}
\usepackage{tcolorbox}
\usepackage{enumitem}
\usepackage[ruled]{algorithm2e}

\usepackage{hyperref}
\hypersetup{hypertex=true,
colorlinks=true,
linkcolor=blue,
anchorcolor=blue,
}

\usepackage{booktabs} 
\usepackage{multirow}
\usepackage{graphicx}
\usepackage{adjustbox}
\usepackage{tabularx}
\newcolumntype{C}[1]{>{\centering\arraybackslash}p{#1}}
\usepackage{colortbl}
\usepackage{graphicx}  
\usepackage{float}  
\usepackage{subfigure}  
\usepackage{color}
\usepackage{arydshln} 
\usepackage{caption}
\usepackage{makecell}

\usepackage{adjustbox}
\usepackage{threeparttable}
\theoremstyle{definition}

\usepackage{microtype}
\newcommand{\ie}[0]{\textit{i.e.,}\xspace}
\newcommand{\eg}[0]{\textit{e.g.,}\xspace}

\newcommand{\todo}[1]{\textcolor{black}{#1}}

\makeatletter

\newcommand{\Rmnum}[1]{\expandafter\@slowromancap\romannumeral #1@}
\makeatother

\newcommand{\tool}{\textsc{AutoMerge}\xspace}

\begin{document}

\title{\tool: Search-Based Model Merging Framework for Effective Model Reuse}


\author{You Lu}
\authornote{Equal contribution.}
\affiliation{%
\institution{Fudan University}
\city{Shanghai}
\country{China}
}

\author{Jiyang Zhang}
\authornotemark[1]
\affiliation{%
\institution{Fudan University}
\city{Shanghai}
\country{China}
}

\author{Bihuan Chen}
\authornote{Bihuan Chen is the corresponding author.}
\affiliation{%
\institution{Fudan University}
\city{Shanghai}
\country{China}
}

\author{Chaofeng Sha}
\affiliation{%
\institution{Fudan University}
\city{Shanghai}
\country{China}
}

\author{Dingji Wang}
\affiliation{%
\institution{Fudan University}
\city{Shanghai}
\country{China}
}

\author{Xin Peng}
\affiliation{%
\institution{Fudan University}
\city{Shanghai}
\country{China}
}


\begin{abstract}
{Software reuse has long been recognized as a critical and widely studied topic in software engineering, offering substantial benefits in reducing development costs, improving software quality, and enhancing operational efficiency. This paradigm extends into deep learning through model reuse. Modern software systems frequently integrate multiple deep learning models, and engineers face challenges in maintaining and reusing them within strict compute and memory constraints. Unlike traditional software reuse, which deals with human-readable code, deep learning model reuse involves high-dimensional parameterized functions whose behavior is only observable empirically and typically requires retraining or fine-tuning. This makes the reuse operation itself a quality-assurance concern. While retraining-based approaches (e.g., fine-tuning, knowledge distillation) have been widely adopted for model reuse, they often suffer from high training costs and capability decay. Moreover, in practical software engineering scenarios, engineers frequently lack access to sufficient computational resources and original training data for multi-task models, further constraining the applicability of retraining-based approaches. Recently, model merging has emerged as a training-free alternative that merges multiple task-specific models with the same architecture without retraining, demonstrating promising results in large language models (LLMs). However, existing model merging techniques are primarily proposed for LLMs using fixed hyperparameters, and have not been systematically evaluated on other model architectures across diverse domains.}

To bridge this gap, we present the first systematic study that evaluates five model merging techniques~on~three distinct model architectures across three domains, \ie NLP, image classification, and autonomous~driving. Our findings reveal that directly applying existing model merging techniques leads to highly inconsistent~results and falls notably short of their success within LLMs. Moreover, a single model merging technique~often~fails to handle the heterogeneous structural properties within a model, limiting its applicability to different model architectures across domains. Furthermore, the effectiveness of model merging techniques is highly sensitive to hyperparameter configurations, thereby constraining their potential for broader adoption.

{Inspired by these insights, we propose \tool, a novel search-based model merging framework that first segments complex models into multiple heterogeneous blocks and then systematically explores the merging space to identify the merging technique and its hyperparameter configuration for each block by Bayesian optimization, producing an outperforming multi-task merged model from multiple source models trained on different tasks. Our evaluation demonstrates that \tool, designed for zero-training settings, improves the average preservation rate (PR) by 23.55\% across all three domains and architectures, and reduces the average preservation discrepancy (PD) by 51.94\% compared with other merging techniques. Besides, our segmentation strategy in \tool improves merging effectiveness by 28.11\%, compared to whole-model hyperparameter search. Further, compared with fine-tuning, \tool cuts the time required to obtain a multi-task model by 62.43\% and the computational cost by 64.34\%, while maintaining competitive capabilities.}

\end{abstract}

\maketitle


\section{Introduction}
\label{sec:intro}

Software reuse~\cite{krueger1992software} has long been recognized as a cornerstone of software engineering, offering~substantial benefits in reducing development costs, improving software quality, and accelerating~the delivery of reliable systems~\cite{chen2024understanding, mohagheghi2007quality, irshad2016capturing}. Decades of research have established software reuse~as~a~widely studied and well-practiced paradigm in software engineering~\cite{frakes2005software,standish2009essay}. With the rapid development of artificial intelligence, the notion of reuse has been extended from traditional software artifacts to deep learning models~\cite{jiang2023empirical,pfeiffer2020adapterhub}. The availability of large-scale pre-trained models on open-source repositories such as HuggingFace~\cite{huggingface} and Kaggle~\cite{Kaggle} has greatly facilitated this process, allowing developers and researchers to leverage pre-trained models rather than training from scratch. Beyond the reuse of a single pre-trained model, a large amount of researches explore how multiple task-specific models can be leveraged as source models to construct a unified multi-task model. This paradigm of model reuse seeks to maximize the preservation of specialized capabilities of source models that have been trained for individual tasks for better generalization.

{Unlike traditional software reuse, which deals with human-readable code, deep learning model reuse involves high-dimensional parameterized functions whose behavior is only observable empirically and typically requires retraining or fine-tuning. This makes the reuse operation itself a quality-assurance concern. In practical software engineering scenarios, engineers frequently lack access to sufficient computational resources and original training data for multi-task models. For instance, in autonomous driving systems (ADS), due to strict privacy regulations, models are often trained independently on isolated data from different countries or cities, creating data silos that prevent effective model integration. The challenge is to integrate these separately trained models into a unified model capable of operating across different environments, while maintaining existing capabilities and reducing engineering and deployment costs. Treating models as components to be merged, systematically exploring configurations to combine task-specific models into a single multi-task model without retraining from scratch, addresses key software engineering challenges in maintaining, evolving, and ensuring the quality of deep learning components.}

\todo{Supplement relevant literature on data isolation and data silos.}

\textbf{Literature.} To preserve the specialized capabilities of different source models when constructing a unified multi-task model, a variety of retraining-based approaches have been proposed, including fine-tuning~\cite{devlin2019bert,hu2022lora}, mixture-of-experts (MoE) \cite{li2025theory}, and knowledge distillation~\cite{thadajarassiri2023knowledge,agand2024knowledge,zhou2025all,gao2024complementary}. Fine-tuning~\cite{dodge2020fine,xu2023improving,tu2025robust} is the most widely adopted strategy, where a pre-trained model is adapted~to~tasks by updating its parameters using the training datasets of the tasks. However, it often incurs~high training cost. MoE approaches~\cite{shen2024efficient,tang2024merging} reduce training cost by routing inputs to different task-specific models, yet they depend on complex expert selection mechanisms and typically enlarge the overall model. Knowledge distillation~\cite{yang2025hyperbolic,cui2025multi,xiang2025dkdm,liu2024small} transfers knowledge from a teacher model to a student model, but it may suffer from capability decay during knowledge transfer.~While~these approaches have substantially advanced reuse of pre-trained models and facilitated multi-task learning, they remain constrained by the capability decay~\cite{waheed2024distill,puigcerver2022adversarial, zhang2023robust} and the high training time consumption~\cite{yao2025pre}, {indicating that effective and efficient model reuse, which can preserve source model capabilities without incurring high training costs or requiring access to original training data, is still an open challenge when constructing a multi-task model from multiple source models.
}
Recently, \textit{model merging} has emerged as a promising approach to obtain multi-task large language models (LLMs) by merging multiple task-specific pre-trained source models with the same~architecture. Within the HuggingFace~\cite{huggingface} platform, more than 38,000 multi-task LLMs have already~been produced using merging techniques, reflecting its rapid adoption in practice. For instance, we~can merge the \textit{WizardMath} model~\cite{luo2025wizardmath} and the \textit{WizardCoder} model~\cite{luo2024wizardcoder} that have been trained~on~different tasks with the \textit{Wizard} model architecture, to obtain a multi-task model that can solve~both mathematical reasoning and programming tasks~\cite{yu2024language}.
Existing model merging techniques can be broadly categorized into weight-based techniques~\cite{utans1996weight,ilharco2022editing,matena2022merging,yang2023adamerging}, which leverage various~rules to formally merge the parameter weights of the source models, and subspace-based techniques~\cite{yadav2023ties,yang2024model,huang2024emr,yu2024language}, which rebuild the models into sparse subspace for reducing conflicts during merging. These~techniques typically involve several hyperparameters, \eg the model weight coefficients (\ie \textit{weight}), and the retention fraction of parameter weights in each source model (\ie \textit{density}). 
Different from fine-tuning, MoE, and knowledge distillation, these techniques merge multiple task-specific models into a unified multi-task model in a training-free approach, representing a significant step forward for model reuse of multiple source~models.

With the growing prominence of Transformer-based LLMs, end-to-end complex models have~become the dominant solutions across diverse domains~\cite{gong2022cmkd,carion2020end,hassani2021escaping,shao2023safety} such as image classification~and autonomous driving. Although model merging techniques with optimal hyperparameter configurations have demonstrated encouraging results within LLMs~\cite{li2023deep,yang2024model,yu2024language}, it remains unclear whether such techniques can be successfully applied to other complex deep learning domains. This gap~highlights the need for a systematic investigation into~the effectiveness of model merging techniques for model reuse of multiple task-specific models with the same architecture across different domains.

\textbf{Empirical Study.} To bridge this gap, we conduct the first empirical study that applies five~state-of-the-art model merging techniques (\ie \textsc{\textsc{Linear}}~\cite{utans1996weight}, \textsc{Task Arithmetic}~\cite{ilharco2022editing},~\textsc{TIES} \cite{yadav2023ties}, \textsc{DARE-Linear}~\cite{yu2024language}, and \textsc{DARE-TIES}~\cite{yu2024language}) across three domains (\ie LLMs, image~classification,~and~auto-nomous driving), and evaluates merged models on multiple tasks to answer two research~questions.
 
\begin{itemize}[leftmargin=*]
    \item \textbf{RQ1 Applicability Analysis.} Can existing model merging techniques be directly applied to different model architectures across diverse domains?
    \item \textbf{RQ2 Sensitivity Analysis.} How sensitive are model merging techniques to different hyperparameter configurations across different model architectures and domains?
\end{itemize}

{In \textbf{RQ1}, our analysis reveals that directly applying existing model merging techniques to image classification and autonomous driving domains leads to highly inconsistent results, and falls~notably short of their success within LLMs. Moreover, a single merging technique is often unable~to~accommodate the heterogeneous structural properties within different parts of a model (e.g., CNN blocks require dense connections while Transformer blocks are inherently sparse), further limiting its applicability across diverse architectures. In \textbf{RQ2}, our analysis indicates that the effectiveness of model merging techniques is highly sensitive to hyperparameter configurations across different models, thereby constraining their potential for broader adoption.}

\textbf{Our Approach.} Motivated by these insights, we propose \tool, a novel search-based model merging framework designed for effective model reuse, which takes multiple task-specific models with the same architecture as the source models and merges them into a multi-task~model.~In particular, to deal with the heterogeneous structural properties, \tool segments the source models into multiple heterogeneous blocks. To search for the optimal merging configuration,~\tool uses Bayesian optimization to automatically identify the merging technique~and its hyperparameter configuration for each block, producing an outperforming multi-task merged model from multiple source models trained on different tasks.

\textbf{Evaluation.}
{We conducted extensive experiments to evaluate the effectiveness and efficiency of \tool. We applied \tool to three distinct model architectures across three domains, \ie \textit{Llama2} architecture~\cite{touvron2023llama}, \textit{CCT} architecture~\cite{hassani2021escaping}, and \textit{Interfuser} architecture~\cite{shao2023safety}. The results indicate that \tool, designed for zero-training settings, improves the average preservation rate (PR) by 23.55\% across all three domains and architectures, and reduces the average preservation discrepancy (PD) by 51.94\% compared with other merging techniques. Our segmentation strategy enhances merging effectiveness by 28.11\%, compared to whole-model hyperparameter search. Compared with fine-tuning, \tool cuts the time required to obtain a multi-task model by 62.43\% and the computational cost by 64.34\%, while maintaining competitive capabilities.}

\textbf{Contribution.} The main contributions of our work are summarized as follows.
\begin{itemize}[leftmargin=*]
\item We conduct a systematic study of model merging techniques across diverse model architectures and domains, revealing critical limitations that hinder their effectiveness for model reuse.
\item We propose \tool, a novel search-based model merging framework that merges multiple task-specific source models to obtain a multi-task model through a training-free approach.
\item We implement a prototype of \tool, and conduct extensive experiments to demonstrate its effectiveness and efficiency for model reuse.
\end{itemize}

\section{Empirical Study}\label{sec:empirical}
To systematically investigate the two research questions introduced in the Sec.~\ref{sec:intro}, we present the first empirical study that applies state-of-the-art model merging techniques across diverse domains, and evaluates the resulting merged models on multiple datasets.

\subsection{Study Design}
\subsubsection{Model Merging Technique Selection}
\label{sec:merge techniques}
Model merging is a rapidly developing direction, with many different merging techniques~\cite{utans1996weight,ilharco2022editing,matena2022merging,yang2023adamerging,yadav2023ties,huang2024emr,yu2024language} being proposed. From these techniques, we select five of the most representative and state-of-the-art techniques as our study subjects across the weight-based (\ie \textsc{Linear}~\cite{utans1996weight} and \textsc{Task Arithmetic}~\cite{ilharco2022editing}) and subspace-based techniques (\ie \textsc{TIES}~\cite{yadav2023ties}, \textsc{DARE-Linear}~\cite{yu2024language} and \textsc{DARE-TIES}~\cite{yu2024language}). 

\textbf{\textsc{Linear}~\cite{utans1996weight}.} This technique directly averages the parameters of neural networks trained on different datasets, making parameter weight averaging an effective strategy for efficient model merging.
By replacing the ensemble of models with a single merged model, it significantly reduces storage and inference costs while retaining comparable capabilities. The hyperparameter of this technique is \textit{weight}, which indicates the contribution of each source model to the merging~process.

\textbf{\textsc{Task Arithmetic}~\cite{ilharco2022editing}.} This technique introduces task vectors that capture task-specific~features and can be applied to other models via simple addition or subtraction of parameter weights. By combining task vectors of different models, the merged model can be adapted to handle multiple tasks, enabling knowledge integration. Similarly, \textit{weight} is the hyperparameter of this technique.

\textbf{\textsc{TIES}~\cite{yadav2023ties}.} This technique identifies two key sources of interference in models for merging, \ie redundant parameters, which provide limited benefit and can harm capability, and sign conflicts, where parameters from different models have opposite signs. By trimming redundant parameters and resolving sign conflicts through sign selection and separated merging, this technique produces a merged model that maintains or even enhances capability. The hyperparameters of this technique include \textit{weight}, and \textit{density} that determines the retention fraction of source models' task vectors.

\textbf{\textsc{DARE-Liner}~\cite{yu2024language}.} This technique combine \textsc{Linear} with DARE, which introduces two preprocessing steps applied to delta parameters before merging, \ie \textit{dropping} and \textit{rescaling}. During \textit{dropping}, a proportion of delta parameters is randomly set to zero, reducing redundancy and the overall parameter count. The \textit{rescaling} step then adjusts the remaining parameters to preserve their contribution. Together, these steps reduce redundant parameters by up to 99\% while maintaining capability, making DARE particularly effective for large-scale language models where redundancy is common. Building on this, \textsc{DARE-Linear} integrates DARE with \textsc{Linear}, combining the efficiency of parameter reduction with the simplicity of weight averaging to achieve effective model merging. \textit{weight} and \textit{density} are the hyperparameters of this technique.

\textbf{\textsc{DARE-TIES}~\cite{yu2024language}.} Similar to \textsc{DARE-Linear}, this technique combines \textsc{TIES} with DARE.~By~first reducing redundancy through dropping and rescaling delta parameters, and then applying trimming, sign selection, and separated merging, DARE-TIES effectively mitigates both parameter redundancy and sign conflicts, achieving more efficient and robust model merging for large-scale language models. The same hyperparameters are used as those in \textsc{DARE-Linear}.

\subsubsection{Source Model Selection.}
We evaluate the effectiveness of the five aforementioned merging techniques across three distinct domains, namely large language models (LLMs), image classification, and autonomous driving. For each domain, we select a representative state-of-the-art model that is both complex and widely recognized, thereby ensuring the generality and rigor of our evaluation.

For the case of \textbf{LLMs}, we employ two publicly available task-specific models from HuggingFace~\cite{huggingface} as the source models, \ie \textit{Llama2-7B-Code}~\cite{llama2_7b_code_liu_p} and \textit{Llama2-7B-Chat}~\cite{llama2-7b-chat-hf}, which are both \textit{Llama2}~\cite{touvron2023llama} architecture. The former is primarily trained on code-related corpora, while the latter focuses on instruction-following tasks. Our objective is to investigate whether their merged models are capable of preserving both coding proficiency and instruction-following capabilities.

For the case of \textbf{image classification}, we employ \textit{CCT-Organism}~\cite{cctOrgnasim} and \textit{CCT-Inanimate}~\cite{cctInanimate} from HuggingFace as the source models, which are implemented using the Compact Convolutional Transformer (\textit{CCT})~\cite{hassani2021escaping} architecture. \textit{CCT} is a hybrid architecture that integrates convolutional feature extraction with transformer-based sequence modeling, thereby achieving competitive capability on image classification. \textit{CCT-Organism} is  capable of recognizing various types of organisms, including both animals and plants. In contrast, \textit{CCT-Inanimate} is designed to recognize a wide range of inanimate objects, such as vehicles, furniture, tools, and electronic devices. Our objective is to investigate whether their merged models can effectively integrate the complementary knowledge and maintain robust image classification capabilities across organism and inanimate classes.

For the case of \textbf{autonomous driving}, we employ the model architecture named \textit{Interfuser}~\cite{shao2023safety}, a representative model architecture in autonomous driving due to its end-to-end design that aligns with the current mainstream trend in the field. Since no publicly available models fine-tuned on distinct driving tasks are suitable for our study, we construct two task-specific source models using the \textit{Carla 8 Towns} training dataset~\cite{shao2023safety}, which contains over 40 million annotated images from eight towns in the CARLA simulator~\cite{Dosovitskiy17}. We partition the training set into two subsets, \ie city scenarios and countryside scenarios, and train \textit{Interfuser} separately on each. Specifically, \textit{Interfuser-City} is trained on the city scenarios, while \textit{Interfuser-Countryside} is trained on the countryside scenarios. Our objective is to evaluate whether the merged models can effectively demonstrate robust driving capabilities across city scenarios and countryside scenarios.

\subsubsection{Benchmark and Metric Selection.}
\label{sec:empirical benchmark}
We select the following benchmarks and metrics to evaluate the capabilities of the source models and their merged models on specific tasks. With respect to \textit{Llama2-7B-Code}, \textit{Llama2-7B-Chat} and their merged models, we adopt the HumanEvalPack~\cite{muennighoff2023octopack}, a benchmark designed to evaluate the coding capabilities of LLMs from two perspectives, namely code explanation and code synthesis, across three representative programming languages, \ie Python, Java, and JavaScript. We employ the widely adopted \textit{Pass@10} metric~\cite{chen2023teaching}, which measures~whether~at least one of the top 10 candidate code snippets generated by the model passes~all~test~cases~and produces correct outputs, to evaluate the capability of coding proficiency. To further access instruction-following capability, we utilize MMLU-Pro~\cite{hendrycks2021measuring,wang2024mmlu}, a comprehensive benchmark of over 50 tasks spanning biology, business, chemistry, science, and economics. These tasks involve complex contexts that require models to interpret instructions accurately and perform multi-step reasoning. We quantify this capability using the \textit{Accuracy} of model responses to benchmark questions.

With respect to \textit{CCT-Organism}, \textit{CCT-Inanimate}, and their merged models, we evaluate the~classification capabilities of these models across diverse image classes by employing the testing dataset of the ImageNet-1k~\cite{russakovsky2015imagenet}. We divide the testing dataset into two subsets that include the organism classes and the inanimate classes, respectively. The capabilities of these models are evaluated on the two subsets using the \textit{Top@1} and \textit{Top@5 Accuracy} metrics~\cite{lapin2015top}, which measure the probability that the ground-truth label appears within the model's top-1 and top-5 predictions, respectively.

With respect to \textit{Interfuser-City}, \textit{Interfuser-Countryside}, and their merged models, we evaluate the driving capabilities of these models across different scenarios via closed-loop testing by integrating them with the CARLA simulator and employing the evaluation dataset of the \textit{CARLA 8 Towns} dataset~\cite{shao2023safety}. We partition the evaluation dataset into two subsets, \ie city scenarios and countryside scenarios, and evaluate the driving capabilities of these models on the two subsets. To measure driving capabilities, we adopt multiple metrics reported by the simulator, including \textit{Route Completion}, which measures the percentage of the route successfully driven, \textit{Infraction Penalty}, which quantifies penalties assigned for traffic violations, and \textit{Driving Score}, which provides an overall performance score by combining route completion rate with the violation penalty~\cite{Dosovitskiy17}.

To further assess the effectiveness of model merging techniques, we introduce two metrics, \ie the \textit{preservation rate} (denoted as $PR$ for short), which measures the extent to which the merged model preserves the capability of the source model on specific task, and the \textit{preservation discrepancy} (denoted as $PD$ for short), which evaluates the variance in preservation across different~tasks. Specifically, given a source model $\mathcal{M}_A$ trained on task $T_A$ and a source model $\mathcal{M}_B$ trained on~task $T_B$, the preservation rate of the merged model $\mathcal{M}_{\text{merged}}$ on tasks $T_A$ and $T_B$ are defined by Eq.~\ref{eq:preservation_rate},
\begin{equation}
    \label{eq:preservation_rate}
    \begin{array}{ll}
    \textit{PR}_{A} =\frac{1}{|\mathcal{E}_A|} \sum_{e \in \mathcal{E}_A} \frac{e{(\mathcal{M}_{\text{merged}})}}{e{(\mathcal{M}_{A})}}, \ 
    \textit{PR}_{B} =\frac{1}{|\mathcal{E}_B|} \sum_{e \in \mathcal{E}_B} \frac{e{(\mathcal{M}_{\text{merged}})}}{e{(\mathcal{M}_{B})}}
\end{array}
\end{equation}
where $\mathcal{E}_A$ and $\mathcal{E}_B$ denote the sets of evaluation metrics corresponding to tasks $T_A$ and $T_B$, respectively, and $e(\cdot)$ returns the evaluation score of a model under metric $e$. Then, the capability preservation discrepancy of $\mathcal{M}_{\text{merged}}$ across $T_A$ and $T_B$ is calculated as $PD = |PR_A - PR_B|$.

\newcommand{\specialcell}[2][c]{%
  \begin{tabular}[#1]{@{}c@{}}#2\end{tabular}
}

\subsection{Applicability Analysis (RQ1)} \label{sec:applicability analysis}
\textbf{Setup.} 
To investigate whether existing model merging techniques designed for LLMs can be directly applied to other domains, \ie image classification and autonomous driving, we adopt the optimal hyperparameter configuration of each merging technique reported in the LLM domain and apply them to models in the other two domains. 
Specifically, for \textsc{Linear} and \textsc{Task Arithmetic}, we set the single hyperparameter \textit{weight} to 0.5. For \textsc{TIES}, \textsc{DARE-Linear}, and \textsc{DARE-TIES}, we set \textit{weight} to 0.5 and \textit{density} to 0.65. The merged models generated by these techniques are denoted as~$\mathcal{M}_{{mt}},$ where $\mathcal{M} \in \{\textit{Llama2}, \textit{CCT}, \textit{Interfuser}\}$ indicates the model architectures in different domains, and $mt \in  \{\textsc{Linear}, \textsc{Task Arithmetic}, \textsc{TIES}, \textsc{DARE-Linear}, \textsc{DARE-TIES}\}$ specifies the merging techniques. We evaluate each merged model on the benchmarks corresponding to its domain and report the metrics, measuring the effectiveness of model merging techniques with the preservation rates of capabilities on two tasks, respectively, and the preservation discrepancy across two tasks.

\begin{table}
    \caption{Effectiveness of Model Merging Techniques on \textit{Llama2} Architecture in LLMs}
    \vspace{-5pt}
    \centering
    \label{tab:llm_results}
    \begin{adjustbox}{width=\textwidth}
    \begin{tabular}{cccccccccccccccc}
      \toprule
      \multirow{2}{*}[-11pt]{\specialcell{Model}} & 
      \multicolumn{6}{c}{\textbf{HumanEvalPack}} & 
      \multicolumn{5}{c}{\textbf{MMLU-Pro}} \\ 
      \cmidrule(lr){2-7} \cmidrule(lr){8-12} 
      & \multicolumn{3}{c}{Code Explanation (Pass@10)} & 
      \multicolumn{3}{c}{Code Synthesis (Pass@10)} & 
      \multicolumn{5}{c}{Instruction Following (Accuracy)} \\ 
      \cmidrule(lr){2-4} \cmidrule(lr){5-7} \cmidrule(lr){8-12} 
      & {\rotatebox[origin=c]{0}{Python $\uparrow$}} 
      & {\rotatebox[origin=c]{0}{Java $\uparrow$}} 
      & {\rotatebox[origin=c]{0}{JavaScript $\uparrow$}} 
      & {\rotatebox[origin=c]{0}{Python $\uparrow$}} 
      & {\rotatebox[origin=c]{0}{Java $\uparrow$}} 
      & {\rotatebox[origin=c]{0}{JavaScript $\uparrow$}} 
      & {\rotatebox[origin=c]{0}{Biology $\uparrow$}} 
      & {\rotatebox[origin=c]{0}{Business $\uparrow$}} 
      & {\rotatebox[origin=c]{0}{Chemistry $\uparrow$}} 
      & {\rotatebox[origin=c]{0}{Science $\uparrow$}} 
      & {\rotatebox[origin=c]{0}{Economics $\uparrow$}} \\ 
      \midrule
      \textit{Llama2-7B-Chat}    & 12.44 & 15.85 & 11.04 & 28.65 & 23.78 & 26.22 & \textbf{43.79} & \textbf{20.28} & \textbf{17.58} & \textbf{24.39} & \textbf{31.75} \\
      \textit{Llama2-7B-Code}    & \textbf{22.50} & \textbf{18.29} & \textbf{16.65} & \textbf{51.41} & \textbf{50.00} & \textbf{45.12} & 29,85 & 18.63 & 12.72 & 18.29 & 27.73 \\
      \hdashline
      \( Llama2_\text{\textsc{Linear}} \)           & 19.21 & 20.06 & 16.46 & 44.51 & 46.95 & 39.63 & 42.26 & 19.39 & 17.67 & \textbf{23.17} & 35.07 \\
      \( Llama2_\text{\textsc{Task Arithmetic}} \)  & 21.52 & \textbf{26.34} & \textbf{18.29} & \textbf{45.12} & \textbf{48.78} & \textbf{42.07} & \textbf{43.65} & \textbf{20.15} & 17.93 & 22.44 & \textbf{35.31} \\
      \( Llama2_\text{\textsc{TIES}} \)             & 19.69 & 14.62 & 13.31 & 42.07 & 26.21 & 38.24 & 30.82 & 17.62 & 16.62 & 22.05 & 27.49 \\
      \( Llama2_\text{\textsc{DARE-Linear}} \)      & 21.52 & \textbf{26.34} & \textbf{18.29} & \textbf{45.12} & 48.70 & \textbf{42.07} & \textbf{43.65} & \textbf{20.15} & 17.93 & 22.44 & \textbf{35.31} \\
      \( Llama2_\text{\textsc{DARE-TIES}} \)       & \textbf{22.01} & 22.93 & 14.57 & 31.09 & 39.02 & 36.59 & 41.42 & 19.14 & \textbf{18.02} & \textbf{23.17} & 33.41 \\
      \bottomrule
    \end{tabular}    
\end{adjustbox} 
\end{table}

\textbf{Results.} Table~\ref{tab:llm_results} indicates that most of the merged \textit{LLama2} models successfully preserve the task-specific capabilities of their source models in both coding and instruction-following tasks. 
$Llama2_\textsc{DARE-Linear}$ performs the best, which achieves a preservation rate ($PR$) of 104.7\% on the coding task and a $PR$ of 101.14\% on the instruction following task, with only 3.51\% preservation discrepancy ($PD$). 
The worst merged model is $Llama2_\textsc{TIES}$, which achieves a $PR$ of 77.73\% on the coding task and a $PR$ of 89.60\% on the instruction following task, with the highest $PD$ of 11.87\%. 

\begin{table}[!t]
  \caption{Effectiveness of Model Merging Techniques on \textit{CCT} Architecture in Image Classification}
  \vspace{-5pt}
  \centering
  \label{tab2:cct_result}
  \begin{adjustbox}{width=0.5\textwidth}
  \begin{tabular}{cccccccccccccccc}
    \toprule
    \multirow{2}{*}{Model}& \multicolumn{2}{c}{Organism Classes} & \multicolumn{2}{c}{Inanimate Classes}\\
    \cmidrule(lr){2-3} \cmidrule(lr){4-5}
     & Top@1 $\uparrow$ & Top@5 $\uparrow$ & Top@1 $\uparrow$ & Top@5 $\uparrow$ \\
    \midrule
    \textit{CCT-Organism} & \textbf{51.48} & \textbf{78.54} & 0.35 & 3.80  \\
    \textit{CCT-Inanimate} & 4.87 & 15.40 & \textbf{40.86} & \textbf{66.53} \\
    \hdashline
    \( CCT_\text{\textsc{Linear}} \) & \textbf{38.02} & \textbf{66.73} & \textbf{32.49} & \textbf{56.88} \\
    \( CCT_\text{\textsc{Task Arithmetic}} \) & 38.01 & 66.72 & 32.48 & 56.88 \\
    \( CCT_\text{\textsc{TIES}} \) &  30.55 & 57.90 & 27.69 & 52.99 \\
    \( CCT_\text{\textsc{DARE-Linear}} \)   & 35.16 & 63.72 & 31.25 & 55.48 \\
    \( CCT_\text{\textsc{DARE-TIES}} \)   & 23.83 & 49.25 & 22.33 & 45.05 \\
    \bottomrule
  \end{tabular}    
  \end{adjustbox}
\end{table}

Table~\ref{tab2:cct_result} shows that most of the merged \textit{CCT} models fail to simultaneously preserve the capabilities of their source models on different tasks. \(CCT_{\textsc{Linear}}\) achieves a $PR$ of 79.41\% on the organism classes, and a $PR$ of 82.51\% on the inanimate classes, with the least $PD$ of 3.10\%. Other merged models are more compromised, with \(CCT_{DARE-TIES}\) exhibiting the most significant capability decay. Specifically, it achieves a $PR$ of only 54.50\% on the organism classes, and a $PR$ of 61.18\% on the inanimate classes, with a $PD$ of 11.87\%, resulting in a near-total loss of image classification capability.

\begin{table}[!t]
  \caption{Effectiveness of Model Merging Techniques on \textit{Interfuser} Architecture in Autonomous Driving}
  \vspace{-5pt}
  \centering
  \label{tab3:ads_result}
  \begin{adjustbox}{width=\textwidth}
  \begin{tabular}{c*{6}{c}}
    \toprule
    \multirow{2}{*}{Model} & 
    \multicolumn{3}{c}{City Scenarios} & 
    \multicolumn{3}{c}{Countryside Scenarios} \\
    \cmidrule(lr){2-4} \cmidrule(lr){5-7}
    & Route Completion $\uparrow$ & Infraction Penalty $\uparrow$ & Driving Score $\uparrow$ & 
     Route Completion $\uparrow$ & Infraction Penalty $\uparrow$ & Driving Score $\uparrow$ \\
    \midrule
    \textit{Interfuser-City}   & \textbf{87.82} & \textbf{0.85} & \textbf{79.60} & 43.46 & 0.51 & 28.43 \\
    \textit{Interfuser-Countryside}   & 36.71 & 0.59 & 25.39 & \textbf{75.68} & \textbf{0.69} & \textbf{42.07} \\
    \hdashline
    \( Interfuser_\textsc{Linear} \)          & 49.60 & 0.59 & 37.45 & 85.39 & 0.65 & 59.87 \\
    \( Interfuser_\textsc{Task Arithmetic} \) & 49.26 & 0.63 & 37.66 & \textbf{87.50} & 0.62 & 57.66 \\
    \( Interfuser_\textsc{TIES} \)            & \textbf{70.62} & \textbf{0.83} & \textbf{63.37} & 51.88 & 0.63 & 33.24 \\
    \( Interfuser_\textsc{DARE-Linear} \)     & 49.45 & 0.59 & 37.43 & 85.74 & \textbf{0.73} & \textbf{65.09} \\
    \( Interfuser_\textsc{DARE-TIES} \)       & 63.57 & 0.75 & 59.73 & 55.69 & 0.56 & 38.92 \\
    \bottomrule
  \end{tabular}    
  \end{adjustbox} 
\end{table}

Similar to the findings in image classification, Table~\ref{tab3:ads_result} shows that none of the merged \textit{Interfuser} models effectively preserve the capabilities of their source models. Although some merged models surpass a source model on certain tasks, they often suffer from substantial capability decay on the other task.
The best-performing merged model is $Interfuser_{\textsc{DARE-TIES}}$, which achieves a $PR$ of 75.03\% in city scenarios and a $PR$ of 92.51\% in countryside scenarios, with a $PD$ of 17.48\%. The worst merged model is $Interfuser_{\textsc{Task Arithmetic}}$, which achieves a $PR$ of 47.01\% in city scenarios and a $PR$ of 154.71\% in countryside scenarios, with the highest $PD$ of 107.70\%.

\begin{tcolorbox}[size=small, opacityfill=0.15, before skip=10pt, after skip=10pt]
  \textit{\textbf{Insight-1.}} When directly applied to image classification and autonomous driving, existing model merging techniques yield high capability preservation discrepancy and fall far short of their success with LLMs, thereby limiting the effectiveness of model reuse to some extent.
\end{tcolorbox}

To further understand the reasons behind the above findings, we conduct a deep analysis of the model architectures and merging techniques. \textsc{Linear} and \textsc{Task Arithmetic} lack mechanisms to resolve sign conflicts, which are frequent in image classification and autonomous driving, causing severe degradation in convolutional layers. \textsc{TIES} alleviates conflicts via \textit{sign selection} and \textit{separated merging}, but may prune critical parameter weights during \textit{trimming}. \textsc{DARE-Linear} and \textsc{DARE-TIES} resolve conflicts by introducing \textit{dropping} and \textit{rescaling}, which may result in incomplete feature representations and unstable capabilities in image classification and autonomous driving. Specifically, we analyze the changes of parameter weights in both the CNN and Transformer blocks within those complex models before and after merging. For the CNN blocks, due to its dense nature, we observe that parameter pruning introduced by \textit{trimming}, or \textit{dropping} and \textit{rescaling} steps leads to over 35\% of the parameter weights being set to zero when the density parameter is set to 0.65 using \textsc{TIES}, \textsc{DARE-Linear} and \textsc{DARE-TIES}. Since CNNs rely heavily on dense connections for effective spatial feature learning, this pruning leads to a substantial loss in feature extraction capability, negatively impacting the merged model. In contrast, the Transformer part of the model, due to its inherent sparsity, \textit{trimming}, or \textit{dropping} and \textit{rescaling} step in these techniques primarily affect already-zero weights, leading to minimal decay of model capability. A single model merging technique with fixed hyperparameters cannot cope with the parameter retention requirements of heterogeneous structural properties in a complex~model.

\begin{tcolorbox}[size=small, opacityfill=0.15, before skip=10pt, after skip=10pt]
  \textit{\textbf{Insight-2.}} A single model merging technique frequently fails to capture the heterogeneous structural properties in complex models across image classification and autonomous driving domains, thereby constraining its applicability to diverse model architectures.
\end{tcolorbox}


\subsection{Sensitivity Analysis (RQ2)}
\textbf{Setup.} To investigate whether the effectiveness of model merging techniques is sensitive to different hyperparameter configurations across different model architectures and domains, we conduct a sensitivity analysis of the \textit{weight} and \textit{density} hyperparameters of different techniques across the three domains. Specifically, for \textsc{Linear} and \textsc{Task Arithmetic}, we vary the \textit{weight} from 0 to 1 with a step size of 0.1. For \textsc{TIES}, \textsc{DARE-Linear}, and \textsc{DARE-TIES}, instead of varying one hyperparameter while fixing the other, we perform a joint grid search over both \textit{weight} and \textit{density}. Specifically, we vary the \textit{weight} from 0.0 to 1.0 with a step size of 0.2, and the \textit{density} from 0.5 to 1.0 with a step size of 0.1, evaluating all possible combinations of the two hyperparameters. We evaluate all the merged models and report their capability preservation rates of their source models on each specific task.

\begin{figure}[!t]
    \centering
    \includegraphics[width=0.98\textwidth]{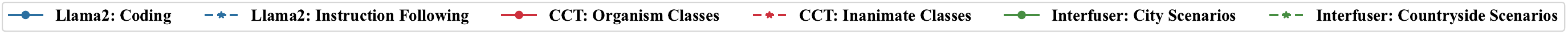} \\[-1ex] 
    
    \subfigure[\textsc{Linear}]{%
        \includegraphics[width=0.35\textwidth]{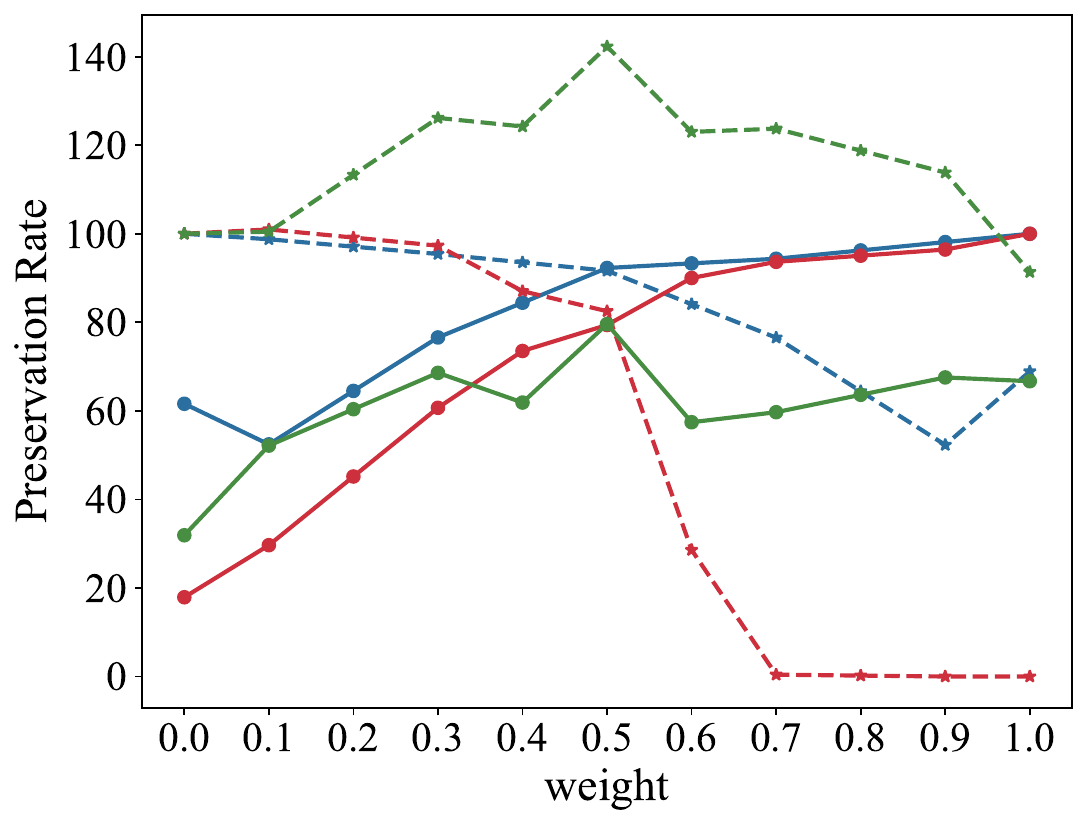}%
        \label{fig:preservation_linear}
    }
    \hspace{0.05\textwidth}
    \subfigure[\textsc{Task Arithmetic}]{%
        \includegraphics[width=0.35\textwidth]{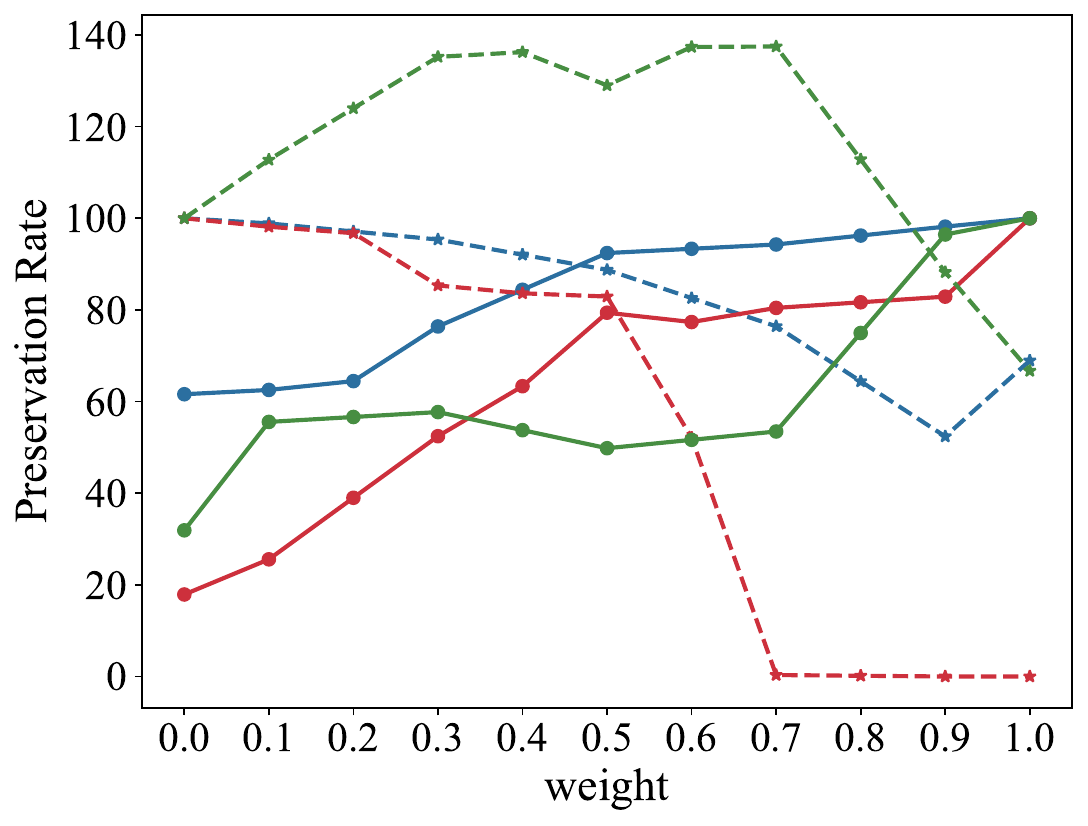}%
        \label{fig:preservation_ta}
    }
    \vspace{-5pt}
    \caption{Weight Sensitivity of \textsc{Linear} and \textsc{Task Arithmetic}}
    \label{fig:weight_linear_TA}
\end{figure}

\textbf{Results.} As shown in Fig.~\ref{fig:weight_linear_TA}, for both \textsc{Linear} and \textsc{Task Arithmetic} techniques, the variation of \textit{weight} has a significant effect on the effectiveness of merged models. 
With respect to the merged \textit{Llama2} models, regardless of the employed techniques, the capability preservation rates ($PR$) of merged models on the coding and instruction-following tasks follow an opposite trend as the value of \textit{weight} increases. The best balance is achieved at \textit{weight} = 0.5, where the merged model preserves average 91.77\% of the both source models' capabilities with the least preservation discrepancy ($PD$) of 2.14\%. 
For the merged \textit{CCT} models, a similar opposite trend can be observed, but the overall effectiveness is notably weaker compared to the results on the \textit{Llama2} architecture. Specifically, when \textit{weight} = 0.5, these two merging techniques perform best. 
With the \textsc{Linear} technique, the merged model achieves an average $PR$ of 80.96\% on two tasks with the least $PD$ of 3.10\%. With the \textsc{Task Arithmetic} technique, the merged model achieves an average $PR$ of 80.94\% on two tasks with the least $PD$ of 3.10\%.
For the merged \textit{Interfuser} models, when \textit{weight} is small, the $PD$ of merged models on the two specific tasks is huge. With the \textsc{Linear} technique, the best result is achieved at \textit{Weight} = 0.5, where the merged model preserves about, on average, 92.18\% of the source models' capabilities with the least $PD$ of 89.76\%. For \textsc{Task Arithmetic}, the optimal point lies~in \textit{weight} = 0.9, where 92.33\% of the source models' capabilities is preserved with the least~$PD$~of~8.23\%. 

\begin{figure}[!t]
    \centering
    
    \subfigure[\textsc{TIES}]{%
        \includegraphics[width=0.3\textwidth]{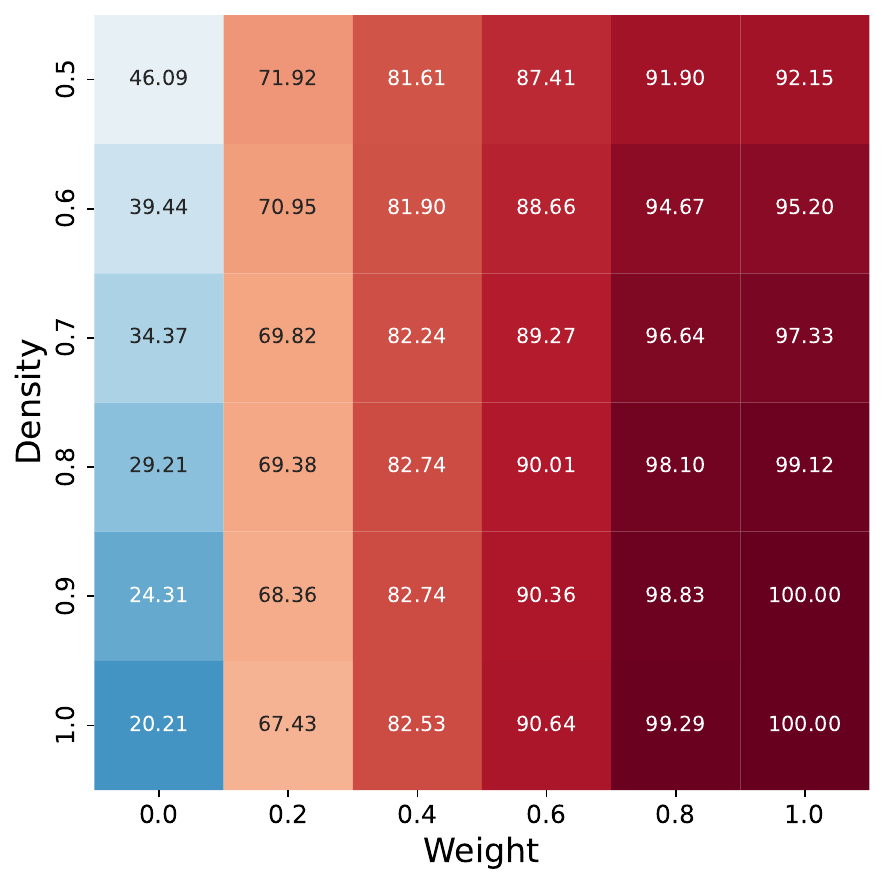}%
        \label{fig:preservation_CCT_ties_taskA}
    }
    \hfill
    \subfigure[\textsc{DARE-Linear}]{%
        \includegraphics[width=0.3\textwidth]{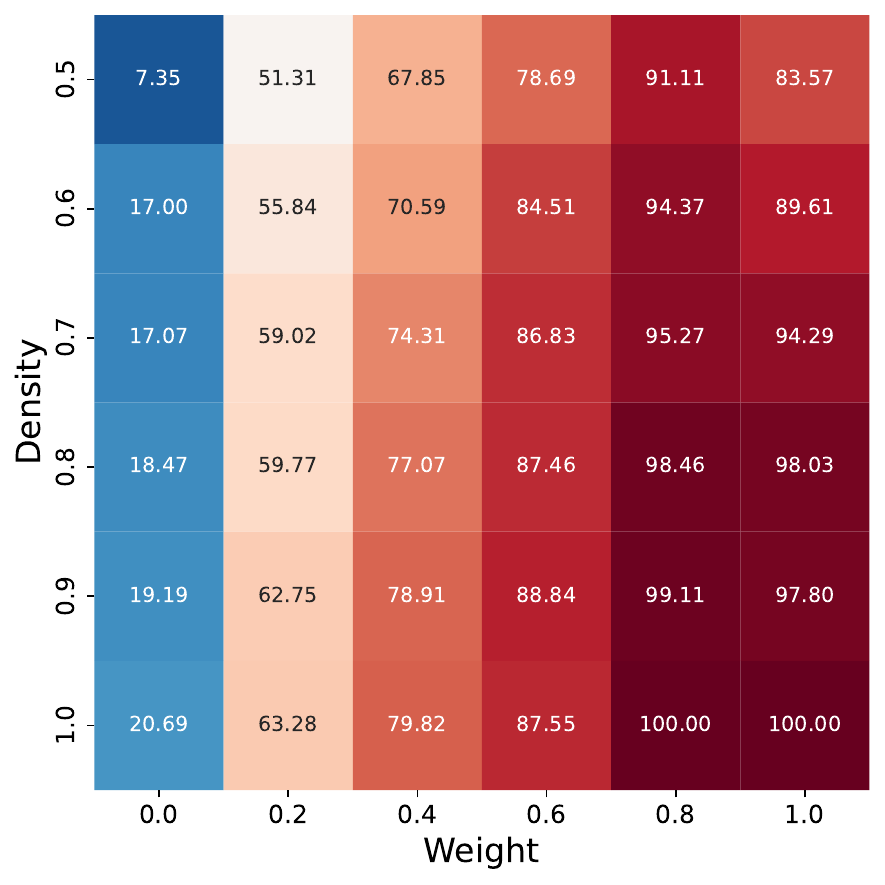}%
        \label{fig:preservation_CCT_dare_linear_taskA}
    }
    \hfill
    \subfigure[\textsc{DARE-TIES}]{%
        \includegraphics[width=0.3\textwidth]{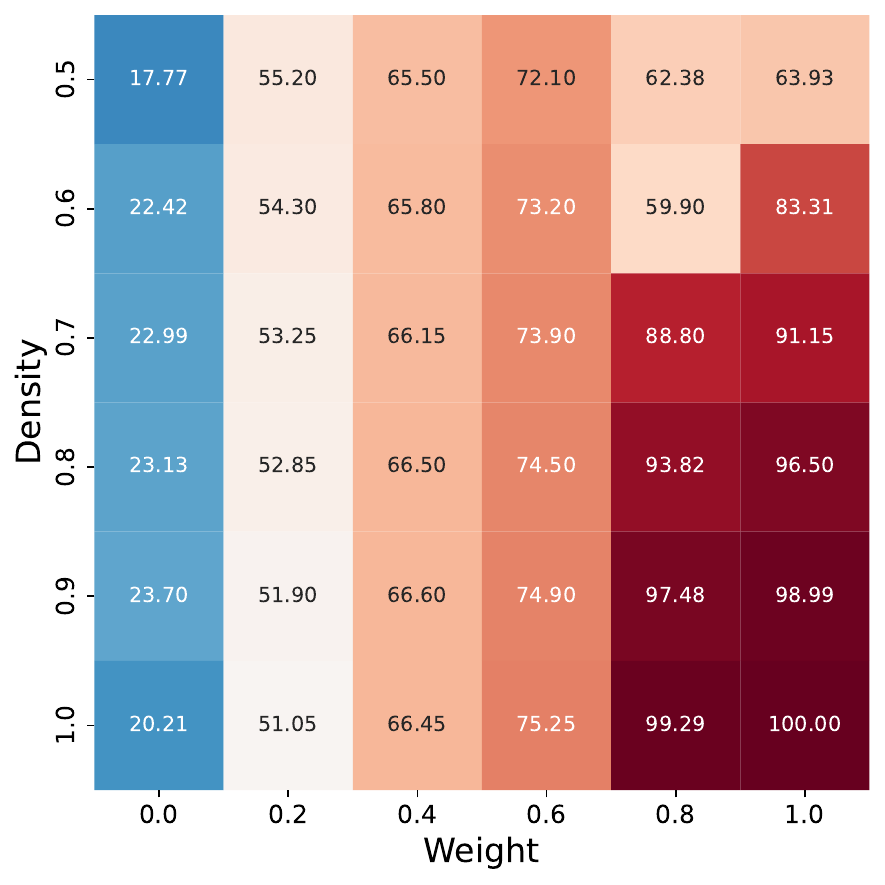}%
        \label{fig:preservation_CCT_dare_ties_taskA}
    }
    \subfigure{%
        \includegraphics[width=0.042\textwidth]{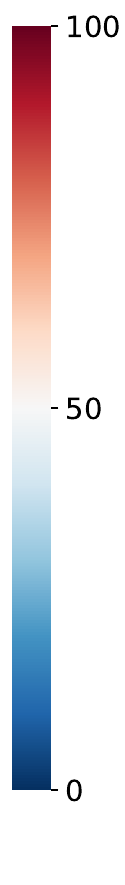}%
    }
    \vspace{-5pt}
    \caption{The Merged CCT Models' Preservation Rate on Organism Classes with Diferent Hyperparameters}
    \label{fig:CCT_taskA_heatmap}
\end{figure}

\begin{figure}[!t]
    \centering
    
    \subfigure[\textsc{TIES}]{%
        \includegraphics[width=0.3\textwidth]{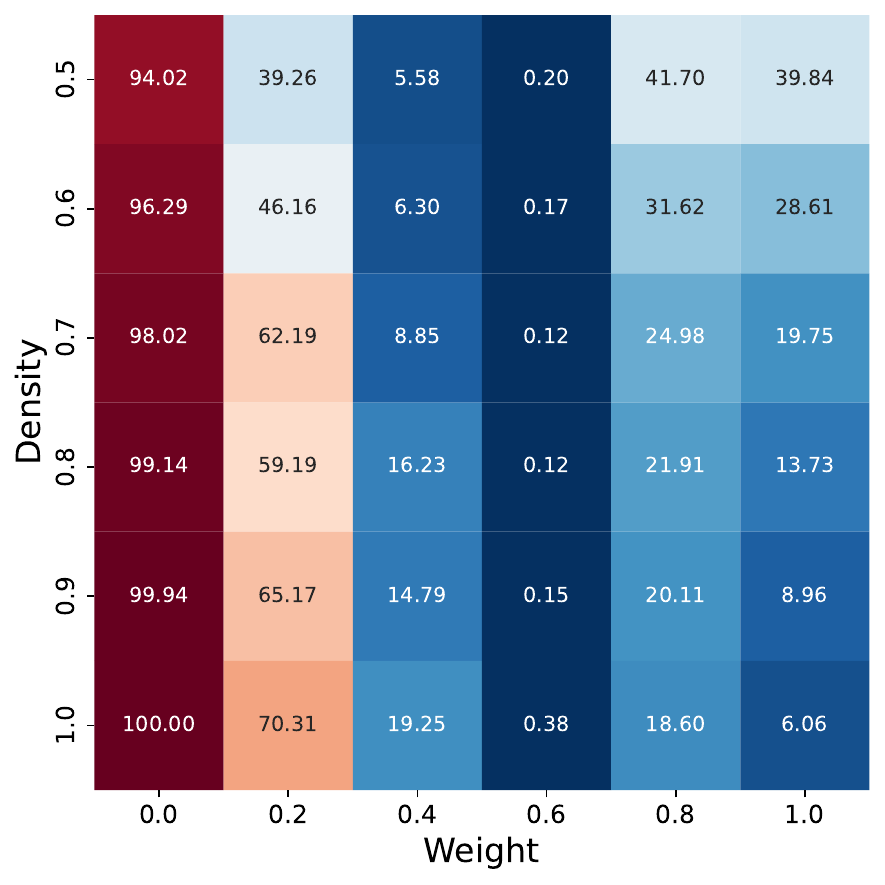}%
        \label{fig:preservation_CCT_ties_taskB}
    }
    \hfill
    \subfigure[\textsc{DARE-Linear}]{%
        \includegraphics[width=0.3\textwidth]{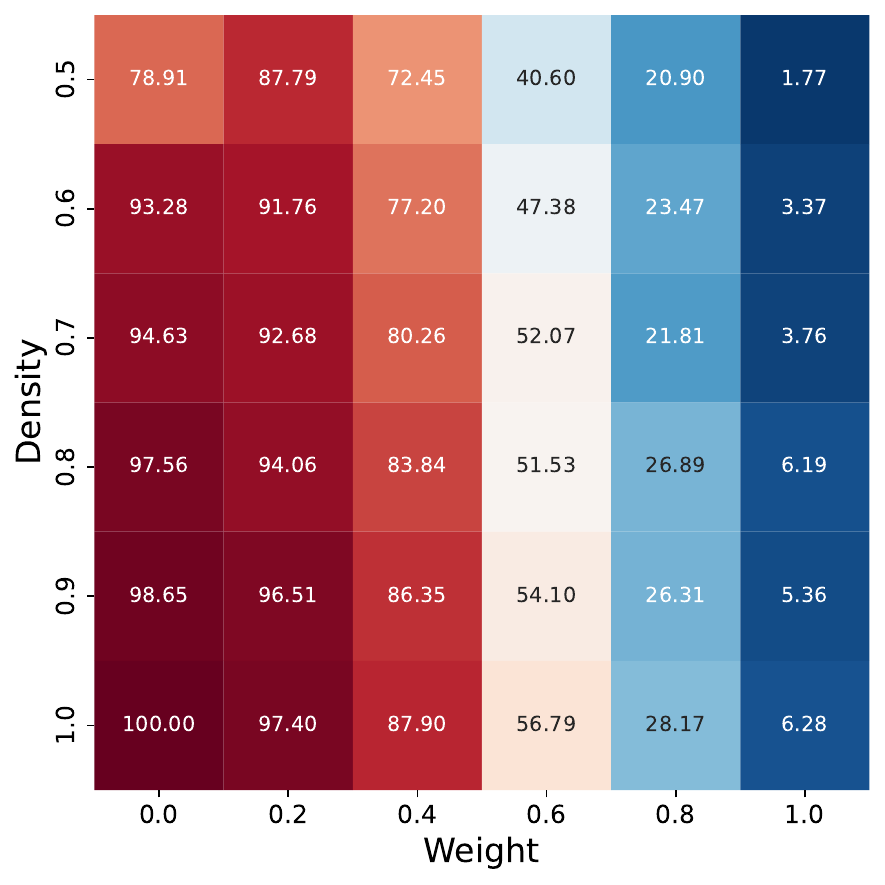}%
        \label{fig:preservation_CCT_dare_linear_taskB}
    }
    \hfill
    \subfigure[\textsc{DARE-TIES}]{%
        \includegraphics[width=0.3\textwidth]{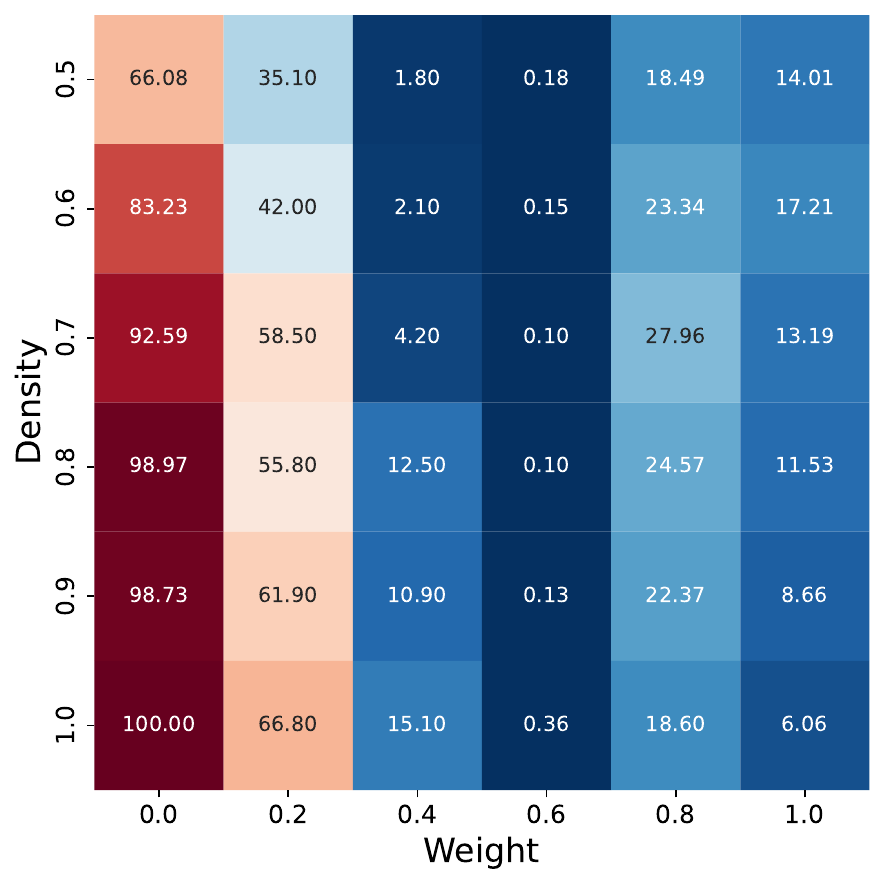}%
        \label{fig:preservation_CCT_dare_ties_taskB}
    }
    \hfill
    \subfigure{%
        \includegraphics[width=0.042\textwidth]{img/heatmap/cbar.pdf}%
    }
    \vspace{-5pt}
    \caption{The Merged CCT Models' Preservation Rate on Inanimate Classes with Diferent Hyperparameters}
    \label{fig:CCT_taskB_heatmap}
\end{figure}

{Figs.~\ref{fig:CCT_taskA_heatmap} and~\ref{fig:CCT_taskB_heatmap} show the PR of merged \textit{CCT} models on organism classes and inanimate classes for \textsc{TIES}, \textsc{DARE-Linear}, and \textsc{DARE-TIES}.
Across all three techniques, the heatmaps exhibit a \textit{systematic trade-off along the \textit{weight} axis}. On organsim classes, PR increases with \textit{weight}: the high-PR region lies in the right half (\textit{weight} 0.6--1.0) and the bottom rows (\textit{density} 0.9--1.0), with peak PR 100\% at (\textit{weight}=1.0, \textit{density}=1.0) for \textsc{TIES} and \textsc{DARE-Linear}. On inanimate classes, PR \textit{decreases} with \textit{weight}: the high-PR region lies in the left half (\textit{weight} 0--0.2), e.g., 100\% at (\textit{weight}=0, \textit{density}=1.0); the right half shows very low PR (e.g., 0.20 at \textit{weight}=0.6, \textit{density}=0.5, and 6.06 at \textit{weight}=1.0, \textit{density}=1.0 for \textsc{TIES}). Thus, the optimal (\textit{weight}, \textit{density}) for organism and inanimate classes do not overlap: improving one task degrades the other. This aligns with the high PD (e.g., 107.70\% for \textsc{Task Arithmetic} in RQ1) and underscores that merging techniques are highly sensitive to \textit{weight} and \textit{density} on the \textit{CCT} architecture, and that a single fixed configuration cannot serve both tasks effectively.}

\begin{figure}[!t]
    \centering
    
    \subfigure[\textsc{TIES}]{%
        \includegraphics[width=0.3\textwidth]{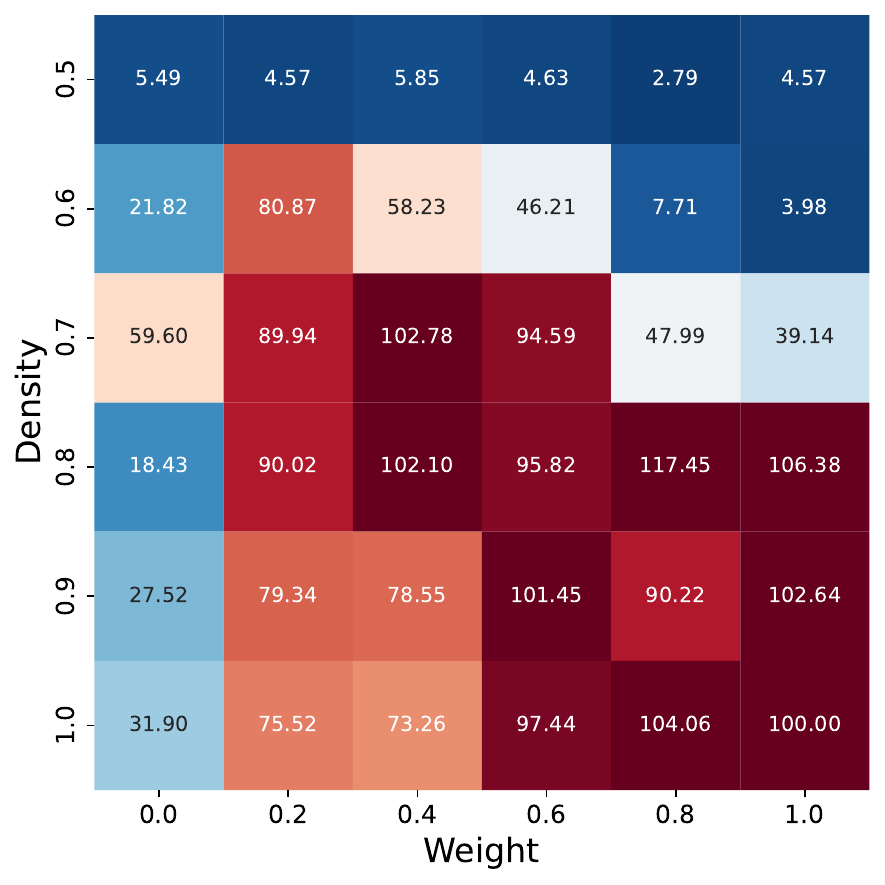}%
        \label{fig:preservation_Interfuser_ties_taskA}
    }
    \hfill
    \subfigure[\textsc{DARE-Linear}]{%
        \includegraphics[width=0.3\textwidth]{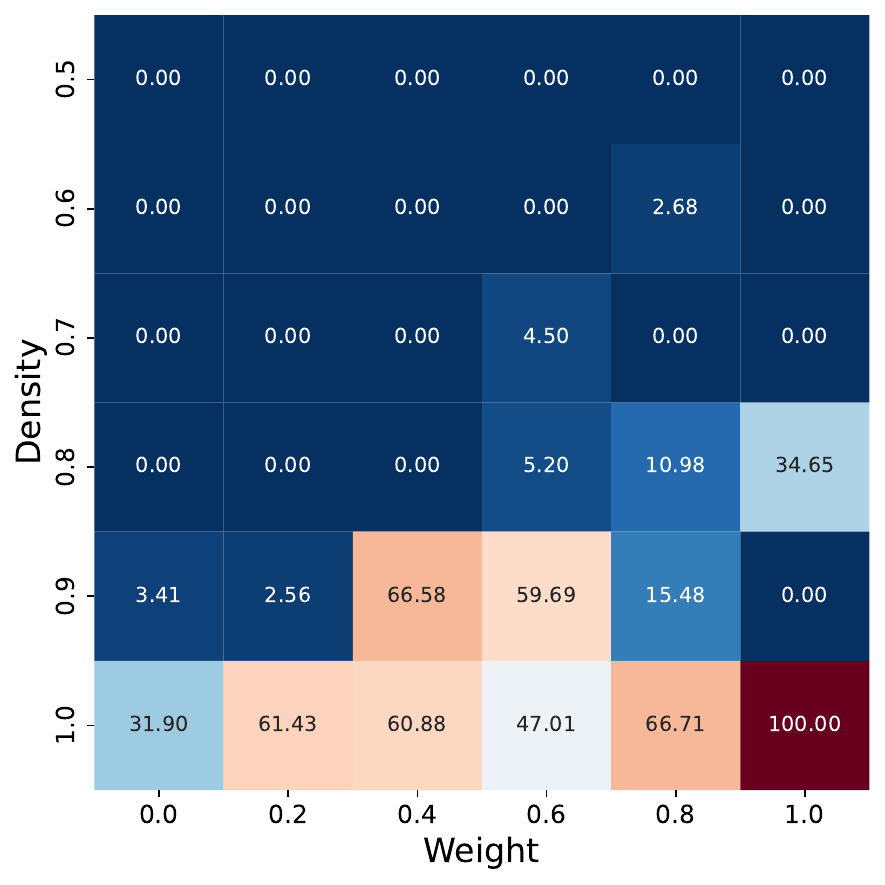}%
        \label{fig:preservation_Interfuser_dare_linear_taskA}
    }
    \hfill
    \subfigure[\textsc{DARE-TIES}]{%
        \includegraphics[width=0.3\textwidth]{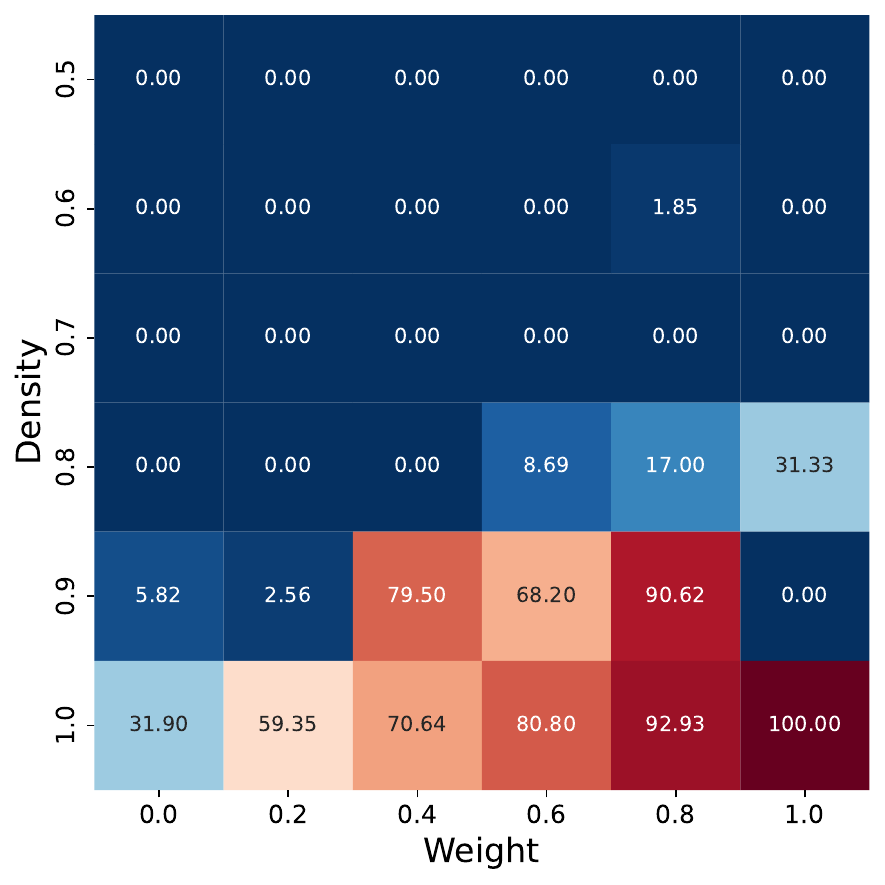}%
        \label{fig:preservation_Interfuser_dare_ties_taskA}
    }
    \hfill
    \subfigure{%
        \includegraphics[width=0.042\textwidth]{img/heatmap/cbar.pdf}%
    }
    \vspace{-5pt}
    \caption{The Merged Interfuser Models' Preservation Rate in City Scenarios with Diferent Hyperparameters}
    \label{fig:Interfuser_taskA_heatmap}
\end{figure}

\begin{figure}[!t]
    \centering
    
    \subfigure[\textsc{TIES}]{%
        \includegraphics[width=0.3\textwidth]{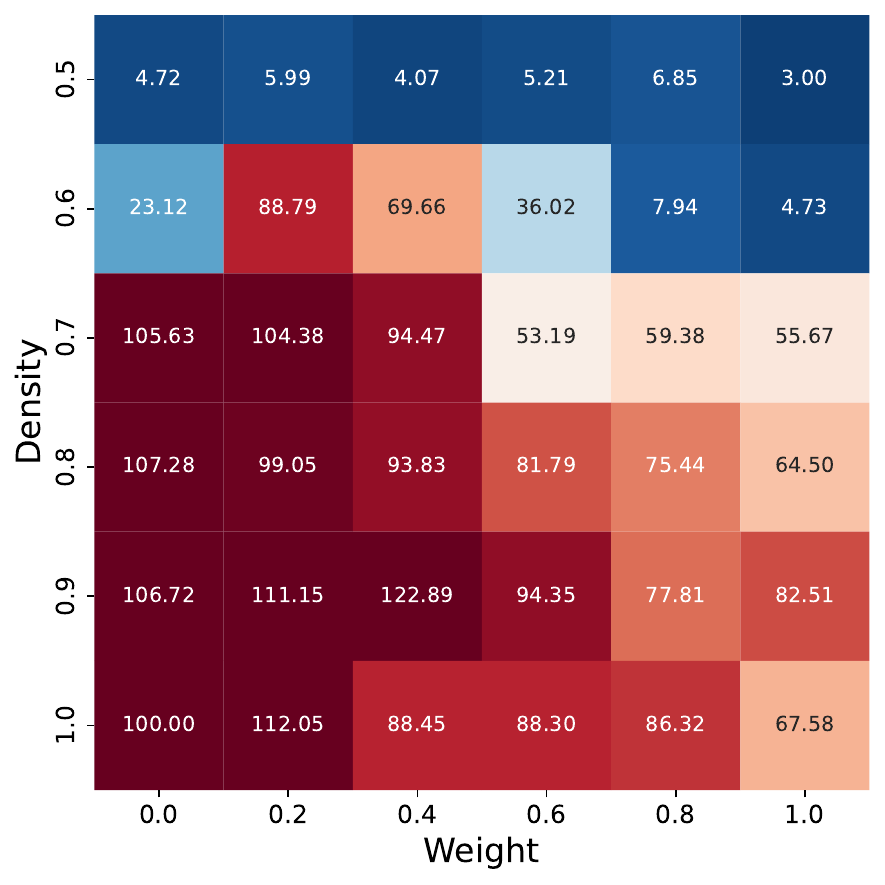}%
        \label{fig:preservation_Interfuser_ties_taskB}
    }
    \hfill
    \subfigure[\textsc{DARE-Linear}]{%
        \includegraphics[width=0.3\textwidth]{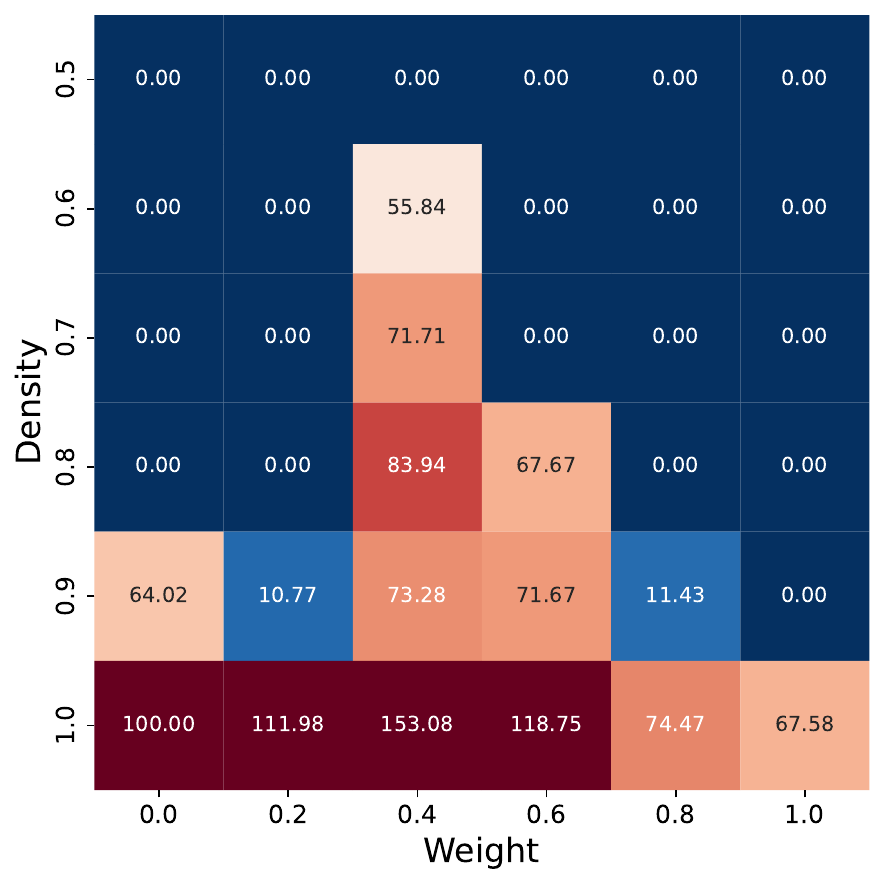}%
        \label{fig:preservation_Interfuser_dare_linear_taskB}
    }
    \hfill
    \subfigure[\textsc{DARE-TIES}]{%
        \includegraphics[width=0.3\textwidth]{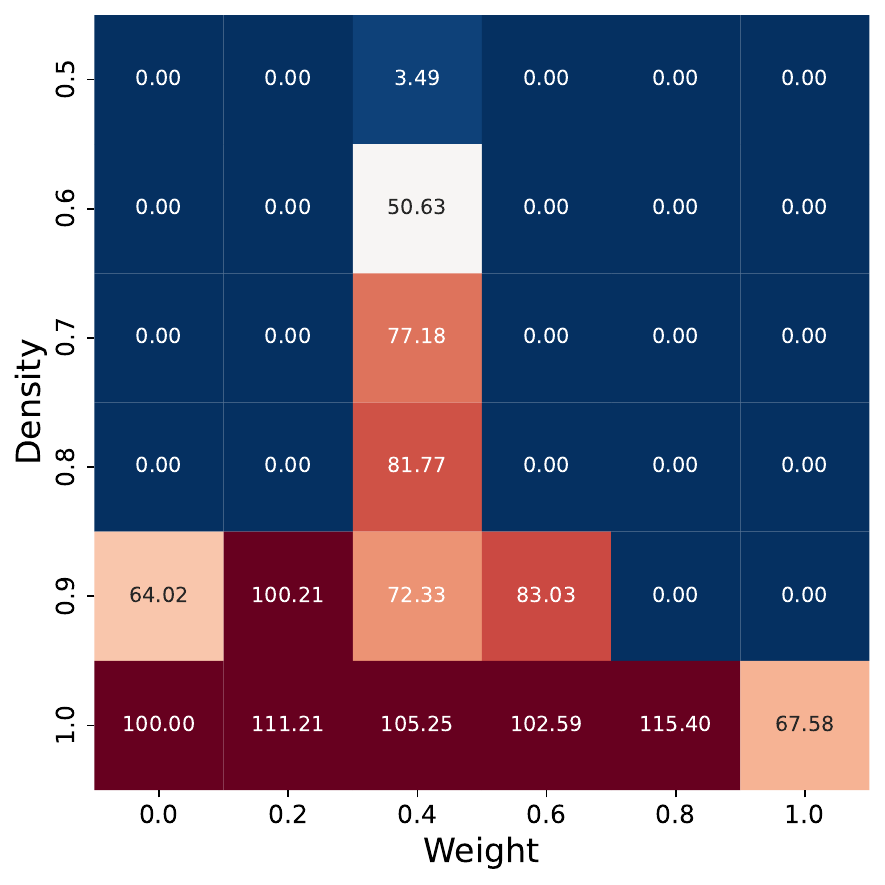}%
        \label{fig:preservation_Interfuser_dare_ties_taskB}
    }
    \hfill
    \subfigure{%
        \includegraphics[width=0.042\textwidth]{img/heatmap/cbar.pdf}%
    }
    \vspace{-5pt}
    \caption{The Merged Interfuser Models' Preservation Rate in Countryside Scenarios with Diferent Hyperparameters}
    \label{fig:Interfuser_taskB_heatmap}
\end{figure}

{Figs.~\ref{fig:Interfuser_taskA_heatmap} and~\ref{fig:Interfuser_taskB_heatmap} show the preservation rate (PR) of merged \textit{Interfuser} models in City Scenarios and Countryside Scenarios over the (\textit{weight}, \textit{density}) grid for \textsc{TIES}, \textsc{DARE-Linear}, and \textsc{DARE-TIES}.
For \textsc{TIES}, when \textit{density} is 0.5, PR remains below 7\% on both tasks across all \textit{weight} values. As \textit{density} increases to 0.7--1.0, high-PR regions emerge, but they lie in \textit{opposite} parts of the \textit{weight} axis: in City Scenarios, the best PR (e.g., 117.45\% at \textit{weight}=0.8, \textit{density}=0.8) occurs at mid-to-high \textit{weight} (0.2--1.0); in Countryside Scenarios, the best PR (e.g., 122.89\% at \textit{weight}=0.4, \textit{density}=0.9) concentrates at low \textit{weight} (0--0.4), and PR drops sharply as \textit{weight} increases. Thus, no single (\textit{weight}, \textit{density}) achieves high PR on both tasks, explaining the high preservation discrepancy (PD) for \textsc{TIES} in RQ1.
For \textsc{DARE-Linear} and \textsc{DARE-TIES}, the heatmaps reveal extreme sensitivity to density. When \textit{density} is 0.5--0.8, PR is zero or near-zero over most of the grid for both tasks; non-negligible PR appears only when \textit{density} $\geq$ 0.9, and strong PR (e.g., $>$80\%) only at \textit{density}=1.0. Moreover, the favorable \textit{weight} again differs between tasks: in City Scenarios, higher \textit{weight} (0.6--1.0) at \textit{density}=1.0 yields PR up to 100\%; on Countryside Scenarios, low \textit{weight} (0--0.4) at \textit{density}=1.0 yields the highest PR (e.g., 153.08\% for \textsc{DARE-Linear} at \textit{weight}=0.4, \textit{density}=1.0). The combination of density-sensitivity and opposite weight preferences makes it impossible to choose one (\textit{weight}, \textit{density}) that works well for both City Scenarios and Countryside Scenarios, consistent with the high PD and the need for block-wise, automated search as in \tool.}

\begin{tcolorbox}[size=small, opacityfill=0.15, before skip=10pt, after skip=10pt]
    \textit{\textbf{Insight-3.}} The effectiveness of model merging techniques is highly sensitive to hyperparameter configurations across different architectures in LLMs, image classification, and autonomous driving domains, thereby constraining their potential for broader adoption in model reuse.
\end{tcolorbox}

\section{Approach}

\begin{figure}[!t]
    \centering
    \includegraphics[width=0.9\textwidth]{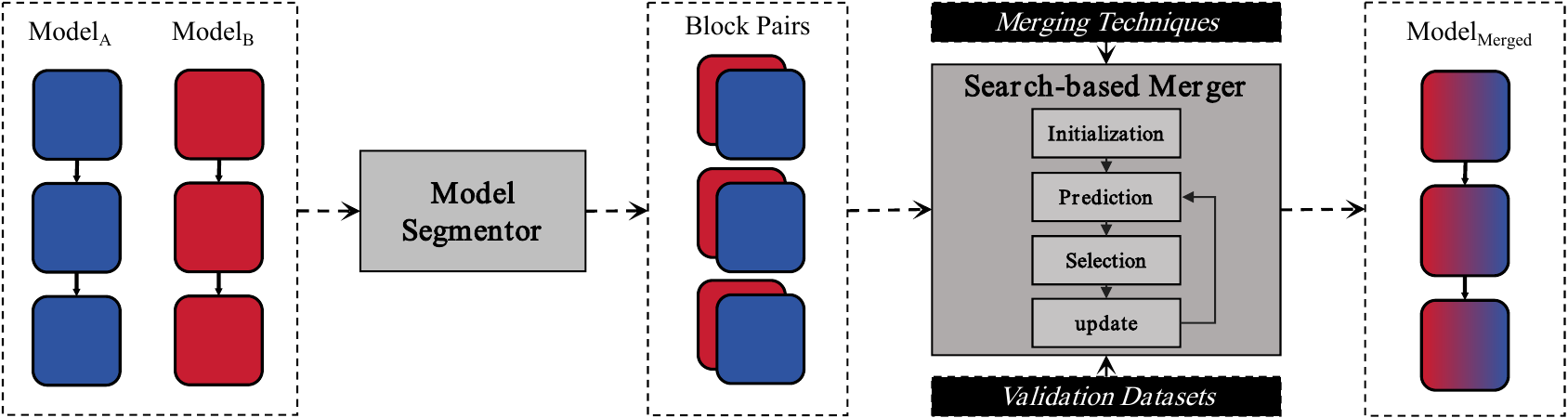}
    \caption{Approach Overview of \tool}
    \label{fig:approach}
    \vspace{-5pt}
\end{figure}

Given \textit{\textbf{Insight-1}}, our goal is to build a model merging framework for
effective model reuse of multiple task-specific models with the same architecture across different domains. Fig.~\ref{fig:approach} shows~the approach overview of our framework, \tool.
Building on \textit{\textbf{Insight-2}}, we propose to apply different merging techniques and hyperparameters to heterogeneous structural properties within complex models. 
For multiple models that share the same architecture but are trained on distinct tasks, we first partition the models into multiple functional blocks based on their intrinsic structural properties through a model segmentor (see Sec.~\ref{sec:segmentor}).
Inspired by \textit{\textbf{Insight-3}}, we adopt a search-based merger (see Sec.~\ref{sec:merger}), which integrates an extendable set of merging techniques and validation datasets of different tasks, to determine the optimal merging configuration for these blocks pairs. After merging, we obtain a new merged model that maximally preserves the capabilities of both source models and minimizes the preservation discrepancy across tasks.
For the ease of presentation, hereafter we illustrate \tool using the case of merging two task-specific models. The same principles, however, generalize naturally to the merging of more than two models.

\subsection{Model Segmentor}\label{sec:segmentor}
A complex model is typically composed of heterogeneous functional blocks, \eg convolutional neural networks (CNNs), Transformers, and multilayer perceptrons (MLPs). These blocks differ in their computational roles, structural characteristics, and data representations. For example, CNNs are well suited for extracting local spatial features, and their intermediate representations usually take the form of four-dimensional tensors $[B, C, H, W]$, where $B$ denotes the batch size, $C$ the number of channels, and $(H, W)$ the spatial resolution. Transformers are designed for modeling long-range dependencies and sequence-level reasoning. Their inputs and outputs are generally three-dimensional tensors $[B, T, D]$, where $T$ is the sequence length and $D$ the embedding dimension. MLPs often appear at the decision-making stage, mapping high-level representations to outputs, with representations typically reduced to two-dimensional tensors~$[B, D]$. Because of these heterogeneous properties, the segmentation of models is guided by shape transitions. Specifically, given the files (\eg $pth$ file in Pytorch~\cite{paszke2017automatic}) storing the parameter weights of a source model, we load the model and record the shape transitions during an inference process to find segmentation points. For instance, the transition from $[B, C, H, W]$ to $[B, T, D]$ marks the shift from CNNs to Transformers, while the reduction from $[B, T, D]$ to $[B, D]$ indicates the~entry~into~an~MLP~layer. 

Formally, given model parameter weights $\Theta$ of model $\mathcal{M}$, the segmentor produces a partition~to generate functional blocks $\{\Theta_1, \Theta_2, \dots, \Theta_K\}$, where $K$ is the number of the blocks.
For a source model $\mathcal{M}_A$ trained on task $\mathcal{T}_A$ and a source model $\mathcal{M}_B$ trained on task $\mathcal{T}_B$, the segmentor outputs a set of aligned block pairs $\{(\Theta^A_k, \Theta^B_k)~|~k \in \left [1, K \right] \}$, which serve as atomic units for merging.~This process ensures that heterogeneous structural properties in the source models are isolated, and enables the search-based merger (see Sec.~\ref{sec:merger}) to apply different merging techniques at the block~level.

\subsection{Search-Based Merger}\label{sec:merger}
Rather than applying a single fixed merging technique across the entire source models, we introduce a search-based merger that automatically selects different merging technique along with its hyperparameters for each block pair generated by the model segmentor. The merged model must strike a balance between high capability preservation of both source models and low preservation discrepancy across tasks.
Consequently, the merging process can be naturally formulated as a multi-objective optimization problem. We begin by specifying the search space and then describe the formulation and solution of this multi-objective optimization problem in detail.

\textbf{Search Space.} For each block pair $(\Theta^A_k, \Theta^B_k)$ of the source models $\mathcal{M}_A$ and $\mathcal{M}_B$, generated by the model segmentor, its search space $\mathcal{S}_k$ can be expressed by Eq.~\ref{eq:block_search_space},
\begin{equation}
    \label{eq:block_search_space}
    \begin{array}{ll}
        \mathcal{S}_k = \{(mt, \textit{weight}, \textit{density})~|~mt \in \mathbb{M}, \textit{weight} \in [0,1], \textit{density} \in (0,1] \}
    \end{array}
\end{equation}
where $mt$ is one in our merging technique set $\mathbb{M}$. We have already integrated five merging techniques into our framework, \ie \textsc{Linear}, \textsc{Task Arithmetic}, \textsc{TIES}, \textsc{DARE-Linear}, and \textsc{DARE-TIES}. Besides, $\textit{weight}$ and $\textit{density}$ are the tunable hyperparameters associated with $mt$, and $\textit{density}$ is empty if $mt \in \{\text{\textsc{Linear}}, \textsc{Task Arithmetic}\}$. 
Thus, the search space for \tool is $ S = \bigcup_{k=1}^K \mathcal{S}_k$.

\textbf{Multi-Objective Optimization.} To solve problem, we adopt the Bayesian optimization~\cite{shahriari2015taking}, a surrogate-based optimization framework well-suited for time-consuming scenarios.~In this framework, a \textit{surrogate model}, \eg a Gaussian process~\cite{williams1995gaussian} or random forest~\cite{hutter2011sequential},~is~constructed~to~provide both predictions of evaluated values and estimates of uncertainty. An \textit{acquisition function} is then~defined on the surrogate model to balance exploration of uncertain regions and exploitation~of~promising areas in the search space, guiding the selection of new candidates. The optimization process often contains four stages, \ie \textit{initialization}, \textit{prediction}, \textit{selection} and \textit{update}, guide by an \textit{objective~function}.

\textit{\textbf{Objective Function.}} As mentioned above, our optimization process has two objectives, \ie maximizing the capability preservation of both source models, and minimizing the preservation discrepancy across tasks. To assess the merged model's capability preservation of the source model on specific task during the optimization process, directly evaluating the metric $PR$ on the testing dataset of the task would be prohibitively time-consuming. To accelerate the search process, we instead approximate capability preservation by comparing the loss values of the merged model and their source models, which share the same loss function, on the integrated validation datasets of different tasks. Formally, given the merged model $\mathcal{M}_{merged}^s$ obtained from the merging configuration sample $s$, and the source models $\mathcal{M}_A$ trained on task $\mathcal{T}_A$ and $\mathcal{M}_B$ trained on task $\mathcal{T}_B$, we define the approximated capability preservation ($AP$) of $\mathcal{M}_{merged}^s$ on $\mathcal{T}_A$ and $\mathcal{T}_B$ by Eq.~\ref{eq:ap},
\begin{equation}
    \label{eq:ap}
    \begin{array}{ll}
        AP_A = \frac{\text{Loss}(\mathcal{M}_A, \mathcal{V}_A)}{Loss(\mathcal{M}_{merged}^s, \mathcal{V}_A)},\ 
        AP_B = \frac{\text{Loss}(\mathcal{M}_B, \mathcal{V}_B)}{Loss(\mathcal{M}_{merged}^s, \mathcal{V}_B)},
    \end{array}
\end{equation}
where $Loss(\cdot)$ returns loss value of a model evaluated on the validation datasets of the corresponding tasks (\ie, $\mathcal{V}_A$ and $\mathcal{V}_B$) {integrated into our framework}. The bigger the $AP$, the higher the capability preservation of the merged model on the corresponding task.

To ensure that the merged model retains the capabilities of both sources without favoring~one task disproportionately, we use the harmonic mean of $AP_A$ and $AP_B$ to construct the final objective function $\mathcal{F} = \frac{2 AP_A \times AP_B}{AP_A + AP_B}$, thus guiding the optimization towards solutions that achieve high multi-task capability preservation and low preservation discrepancy across tasks.

\textit{\textbf{Initialization.}}
In the initialization stage, we first generate 20 merging configuration samples in the search space of block pairs. These samples serve as the starting points for the optimization process. To ensure that the initial samples provide a representative and well-spread coverage of the space, we adopt the Sobol sequence~\cite{sobol1998quasi}, a quasi-random low-discrepancy sampling strategy. Compared with purely random initialization, the Sobol approach mitigates the risk of clustering samples in a narrow region and increases the diversity of observations, thereby strengthening the basis for balancing exploration and exploitation in subsequent iterations.

Each merged model generated from each initial configuration sample $s$ is then evaluated on the objective function $\mathcal{F}$, yielding corresponding objective values. For the surrogate model denoted~as $f(s)$, we adopt a random forest~\cite{hutter2011sequential}, which is better suited than Gaussian processes~\cite{williams1995gaussian} for handling high-dimensional and complex search spaces~\cite{snoek2012practical}. By training the surrogate model on the objective values of these initial samples, the Bayesian optimization process is bootstrapped with a reliable starting model that can subsequently guide prediction and selection more effectively. Among the 20 initial configuration samples, the first 19 samples are used to train the surrogate model, while the 20th configuration sample serves as the starting point for the optimization loop, denoted as $s_1$. After this initialization, the candidate configuration sample at the iteration $t$ is denoted as $s_{t}$, and the next configuration sample $s_{t+1}$ is chosen based on the surrogate model's predictions.

\textit{\textbf{Prediction.}}
At each iteration \( t \), the surrogate model computes the predictive mean \( \hat{f}(s) \) and the uncertainty value \( \sigma(s) \) for each configuration $s$ sampled from iterations 1 to \( t \). The predictive mean \( \hat{f}(s) \) represents the estimated objective value, while the uncertainty value \( \sigma(s) \) is derived from the variance among the ensemble of trees in the random forest. The set \( D_{1:t} = \{ \langle s_i, \hat{f}(s_i), \sigma(s_i) \rangle \}_{i=1}^t \) denotes the collection of configuration samples \( s_i \), their corresponding predictive means \( \hat{f}(s_i) \), and uncertainty values \( \sigma(s_i) \) across iterations 1 to \( t \).

\textit{\textbf{Selection.}}
Based on the predictive mean $\hat{f}(s)$ and uncertainty value $\sigma(s)$ provided by the surrogate model from the previous $t$ iterations, we guide the selection of next configuration sample $s_{t+1}$ with an acquisition function. Specifically, we adopt the Log Expected Improvement (LEI) criterion~\cite{hutter2009experimental}, a logarithmic variant of the classical Expected Improvement (EI)~\cite{jones1998efficient}, as the acquisition function. The classical EI may overemphasize extreme predictions, leading to overexploitation of isolated samples. LEI addresses this by applying a logarithmic transformation to the improvement term, smoothing outlier influence, and yielding a more stable balance between exploration and exploitation. At each iteration $t$, the next sample $s_{t+1}$ is selected by maximizing the LEI over the surrogate model, \ie $s_{t+1} = \arg \max_{s \in S} \text{LEI}(s; D_{1:t})$, exploring promising configurations to yield optimal objective values.


\textit{\textbf{Update.}} In the update stage, the model merged by the configuration sample $s_{t}$ is evaluated~on the objective function $\mathcal{F}$. The obtained objective value is then appended to the training set, together with all previously objective values of samples. With this enlarged dataset, the surrogate model $f(s)$ is retrained to approximate the objective function. In this way, the surrogate refines its approximation of the objective function as more evaluations are accumulated. This update step ensures that the Bayesian optimization process dynamically incorporates the most recent performance feedback, thereby improving the reliability of subsequent predictions and selections.

The prediction-selection-update procedure is repeated in each iteration until the predefined maximum number of iterations is reached. Through this process, the search-based merger ultimately outputs an optimal merging configuration where the merged model achieves high capability~preservation and low preservation discrepancy across tasks.


\section{Evaluation}
We implement a prototype of \tool with 3,256 lines of python code. To evaluate the effectiveness and efficiency of \tool, we design four research questions as follows.

\begin{itemize}[leftmargin=*]
  \item \textbf{RQ3 Effectiveness Evaluation.} What is the effectiveness of \tool in terms of capability preservation of both source models, compared to previous model merging techniques?
  \item \textbf{RQ4 Efficiency Evaluation.} What is the efficiency of \tool in terms of time consumption and computational resource usage?
  \item \textbf{RQ5 Ablation Study.} What is the contribution of our model segmentor to the effectiveness and efficiency of \tool?
  \item \textbf{RQ6 Practical Applicability.} How effective and efficient is \tool for model reuse,~compared to retraining-based approaches?

\end{itemize}

\subsection{Experiment Setup}

For \textbf{RQ3} and \textbf{RQ4}, we apply \tool to the task-specific source models of different architectures used in Sec.~\ref{sec:empirical} across the LLMs, image classification, and autonomous driving domains, respectively, achieving merged models, namely $\textit{Llama2}_{\tool}$, $\textit{CCT}_{\tool}$, and $\textit{Interfuser}_{\tool}$. For \textbf{RQ3}, we first evaluate the merged models obtained by \tool on the benchmarks corresponding to their domains, and report the metrics, comparing the capabilities of these merged models with their source models across different tasks. Then, we compare the capability preservation rate and preservation discrepancy across tasks of the merged models obtained by \tool with those of the merged models obtained by five state-of-the-art model merging techniques,~\ie \textsc{Linear}, \textsc{Task Arithmetic}, \textsc{TIES}, \textsc{DARE-Linear}, and \textsc{DARE-TIES}. For \textbf{RQ4}, we record the average time consumption, peak GPU memory usage, and GPU utilization of \tool during merging.

For \textbf{RQ5}, we design a variant of \tool that removes the model segmentor, denoted~as~\textit{w/o segmentor}, to merge the source models in three domains, which searches the optimal merging~technique and its hyperparameters for the entire model as a whole. With respect to effectiveness, we evaluate the capabilities of models merged by this variant on specific tasks, and compare them with corresponding merged models obtained by \tool. With respect to efficiency, we compare the time consumption and computational resource usage of \tool and this variant.

For \textbf{RQ6}, we fully fine-tune the \textit{Llama2-7B-Chat} on the coding training dataset in CodeSearchNet~\cite{husain2019codesearchnet} for 4 epochs to obtain the model $\textit{Llama2}_{\text{training}}$. \textit{CCT-Organism} is fully fine-tuned on the inanimate classes of the ImageNet-1K data~\cite{russakovsky2015imagenet} for 300 epochs to obtain the model $\textit{CCT}_{\text{training}}$. 
\textit{Interfuser-City} is fully fine-tuned on the training dataset of countryside scenarios in \textit{CARLA 8 Towns} dataset~\cite{shao2023safety} for 200 epochs, obtaining the model $\textit{Interfuser}_{\text{training}}$. 
We then compare the capability of the merged models obtained by \tool with that of the fully fine-tuned models, and report their time consumption and computational resource usage.

\textbf{Experiment Environment,} We conduct all the experiments on Ubuntu 20.04.4 LTS servers with 4 NVIDIA GeForce RTX 3090 GPUs, Intel(R) Xeon(R) Silver 4310 @ 2.10GHz and 128GB memory.


\subsection{Effectiveness Evaluation (RQ3)}

\begin{table}[!t]
    \caption{Effectiveness of \tool on \textit{Llama2} Architecture in LLMs}
    \vspace{-5pt}
    \centering
    \label{tab:llm_baymerge_results}
    \begin{adjustbox}{width=\textwidth}
    \begin{tabular}{cccccccccccccccc}
      \toprule
      \multirow{2}{*}[-11pt]{\specialcell{Model}} & 
      \multicolumn{6}{c}{\textbf{HumanEvalPack}} & 
      \multicolumn{5}{c}{\textbf{MMLU-Pro}} \\ 
      \cmidrule(lr){2-7} \cmidrule(lr){8-12} 
      & \multicolumn{3}{c}{Code Explanation (Pass@10)} & 
      \multicolumn{3}{c}{Code Synthesis (Pass@10)} & 
      \multicolumn{5}{c}{Instruction Following (Accuracy)} \\ 
      \cmidrule(lr){2-4} \cmidrule(lr){5-7} \cmidrule(lr){8-12} 
      & {\rotatebox[origin=c]{0}{Python $\uparrow$}} 
      & {\rotatebox[origin=c]{0}{Java $\uparrow$}} 
      & {\rotatebox[origin=c]{0}{JavaScript $\uparrow$}} 
      & {\rotatebox[origin=c]{0}{Python $\uparrow$}} 
      & {\rotatebox[origin=c]{0}{Java $\uparrow$}} 
      & {\rotatebox[origin=c]{0}{JavaScript $\uparrow$}} 
      & {\rotatebox[origin=c]{0}{Biology $\uparrow$}} 
      & {\rotatebox[origin=c]{0}{Business $\uparrow$}} 
      & {\rotatebox[origin=c]{0}{Chemistry $\uparrow$}} 
      & {\rotatebox[origin=c]{0}{Science $\uparrow$}} 
      & {\rotatebox[origin=c]{0}{Economics $\uparrow$}} \\ 
      \midrule
      \textit{Llama2-7B-Chat}    & 12.44 & 15.85 & 11.04 & 28.65 & 23.78 & 26.22 & \textbf{43.79} & \textbf{20.28} & \textbf{17.58} & \textbf{24.39} & \textbf{31.75} \\
      \textit{Llama2-7B-Code}    & \textbf{22.50} & \textbf{18.29} & \textbf{16.65} & \textbf{51.41} & \textbf{50.00} & \textbf{45.12} & 29,85 & 18.63 & 12.72 & 18.29 & 27.73 \\
      \hdashline
      \( Llama2_\text{\textsc{\tool}} \)  & 22.43 & \textbf{26.34} & \textbf{18.29} & 45.12 & 48.78 & 42.03 & 42.26 & \textbf{23.31} & \textbf{17.67} & 23.17 & \textbf{35.31} \\
      \bottomrule
    \end{tabular}    
\end{adjustbox} 
\end{table}

\begin{table}[!t]
    \caption{Effectiveness of \tool on \textit{CCT} Architecture in Image Classification}
    \vspace{-5pt}
    \centering
    \label{tab:cct_baymerge_result}
    \begin{adjustbox}{width=0.5\textwidth}
    \begin{tabular}{cccccccccccccccc}
      \toprule
      \multirow{2}{*}{Model}& \multicolumn{2}{c}{Organism Classes} & \multicolumn{2}{c}{Inanimate Classes}\\
      \cmidrule(lr){2-3} \cmidrule(lr){4-5}
       & Top@1 $\uparrow$ & Top@5 $\uparrow$ & Top@1 $\uparrow$ & Top@5  $\uparrow$\\
      \midrule
      \textit{CCT-Organism} & \textbf{51.48} & \textbf{78.54} & 0.35 & 3.80  \\
      \textit{CCT-Inanimate} & 4.87 & 15.40 & \textbf{40.86} & \textbf{66.53} \\
      \hdashline
      \(CCT_{\textsc{\tool}}\) & 47.36 & 70.45 & 34.71 & 56.76 \\
      \bottomrule
    \end{tabular}    
    \end{adjustbox}
\end{table}

With respect to the effectiveness of \tool on the \textit{Llama2} architecture in LLMs domain, Table~\ref{tab:llm_baymerge_results} shows that $Llama2_{\textsc{\tool}}$ achieves a $PR$ of 105.33\% on the coding task and a $PR$ of 103.63\% on the instruction-following task, respectively, and the $PD$ of $Llama2_{\textsc{\tool}}$ between coding and instruction-following task is 1.70\%. Notably, $Llama2_{\textsc{\tool}}$ surpasses \textit{Llama2-7B-Code} in explaining Java and JavaScript code, and outperforms \textit{Llama2-7B-Chat} on questions in business, chemistry, and economics, indicating that \tool can preserve or even strengthen the capabilities of the source models in the merged model on specific tasks.
\begin{table}[!t]
  \caption{Effectiveness of \tool on \textit{Interfuser} Architecture in Autonomous Driving}
  \vspace{-5pt}
  \centering
  \label{tab:ads_baymerge_results}
  \begin{adjustbox}{width=\textwidth}
  \begin{tabular}{c*{6}{c}}
    \toprule
    \multirow{2}{*}{Model} & 
    \multicolumn{3}{c}{City Scenarios} & 
    \multicolumn{3}{c}{Countryside Scenarios} \\
    \cmidrule(lr){2-4} \cmidrule(lr){5-7}
    & Route Completion $\uparrow$ & Infraction Penalty $\uparrow$ & Driving Score $\uparrow$ & 
    Route Completion $\uparrow$ & Infraction Penalty $\uparrow$ & Driving Score $\uparrow$ \\
    \midrule
    \textit{Interfuser-City}   & \textbf{87.82} & \textbf{0.85} & \textbf{79.60} & 43.46 & 0.51 & 28.43 \\
    \textit{Interfuser-Countryside}  & 36.71 & 0.59 & 25.39 & \textbf{75.68} & \textbf{0.69} & 42.07 \\
    \hdashline
    \(Interfuser_{\textsc{\tool}}\) & \textbf{96.00} & \textbf{1.09} & \textbf{89.51} & \textbf{86.72} & \textbf{0.78} & \textbf{47.83} \\
    \bottomrule
  \end{tabular}    
  \end{adjustbox} 
\end{table}
With respect to the effectiveness of \tool on the \textit{CCT} architecture in image classification domain, Table~\ref{tab:cct_baymerge_result}~shows that $CCT_{\textsc{\tool}}$ achieves a $PR$ of 90.84\% on the organism classes and a $PR$ of 85.11\% on inanimate classes with a $PD$ of 5.73\%. 
\begin{figure}[!t]
  \centering
  \includegraphics[width=0.98\textwidth]{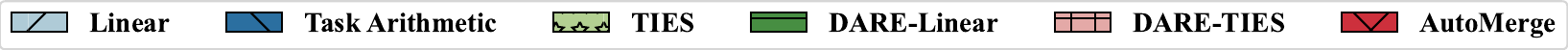} \\[-1ex] 
  
  \subfigure[Llama2]{%
      \includegraphics[width=0.3\textwidth]{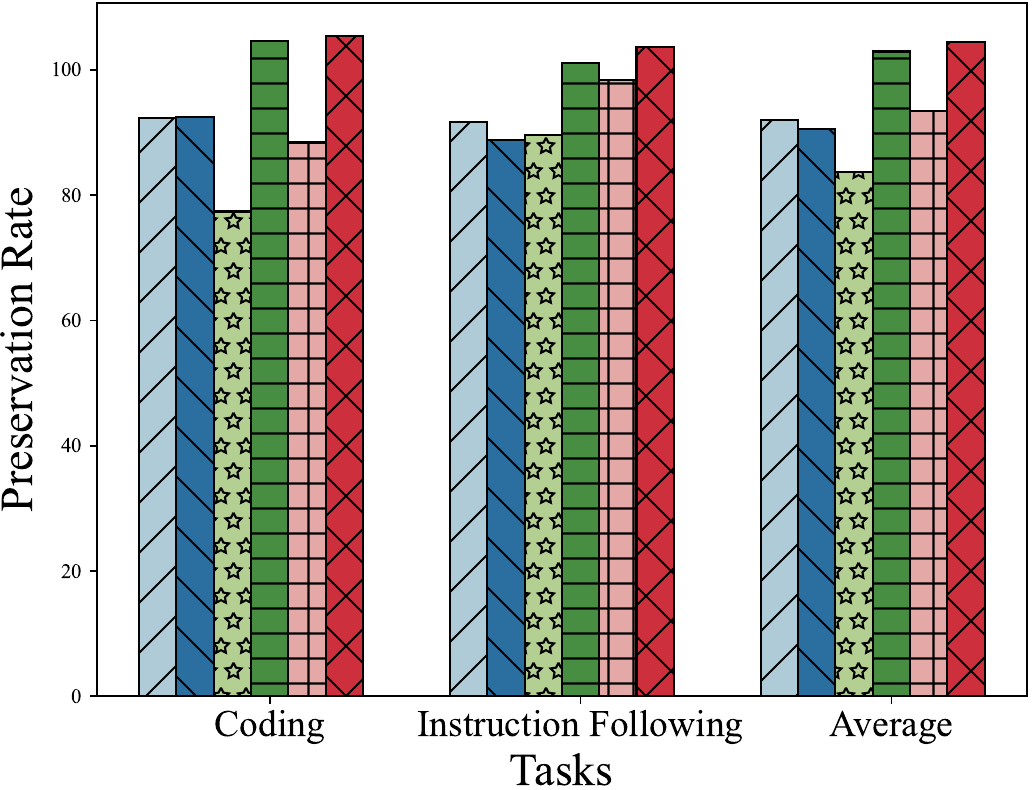}%
      \label{fig:preservation_Llama}
  }
  \hspace{0.02\textwidth}
  \subfigure[CCT]{%
      \includegraphics[width=0.3\textwidth]{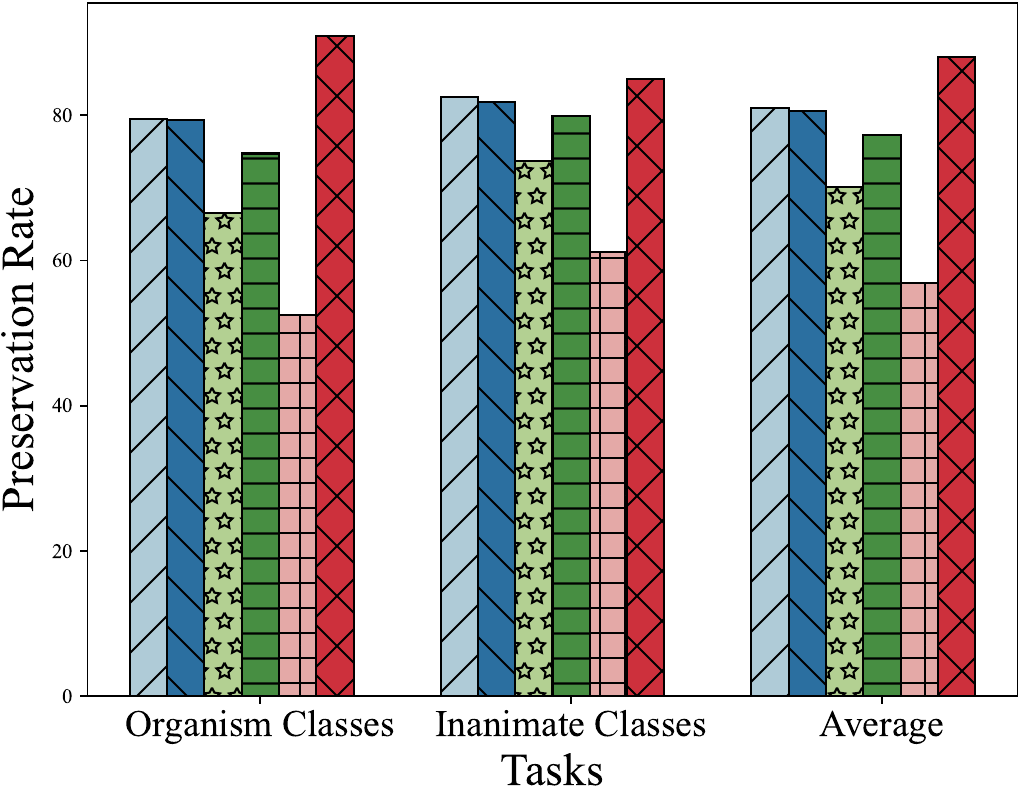}%
      \label{fig:preservation_CCT}
  }
  \hspace{0.02\textwidth}
  \subfigure[Interfuser]{%
      \includegraphics[width=0.3\textwidth]{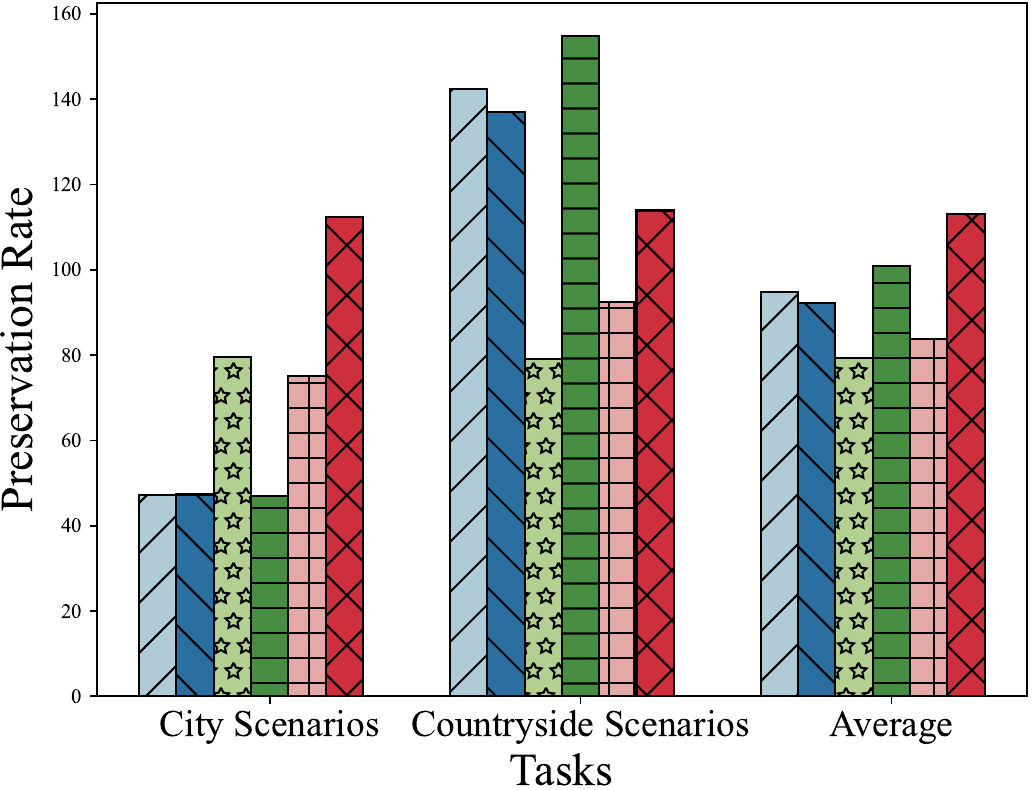}%
      \label{fig:preservation_Interfuser}
  }
  \vspace{-5pt}
  \caption{Capability Preservation of Source Models across Tasks}
  \label{fig:comparision_effectiveness}
\end{figure}
\begin{figure}[!t]
  \centering
  \includegraphics[width=0.5\textwidth]{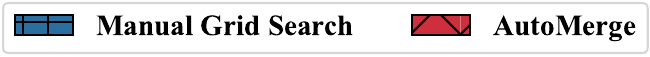} \\[-1ex] 
  
  \subfigure[Llama2]{%
      \includegraphics[width=0.3\textwidth]{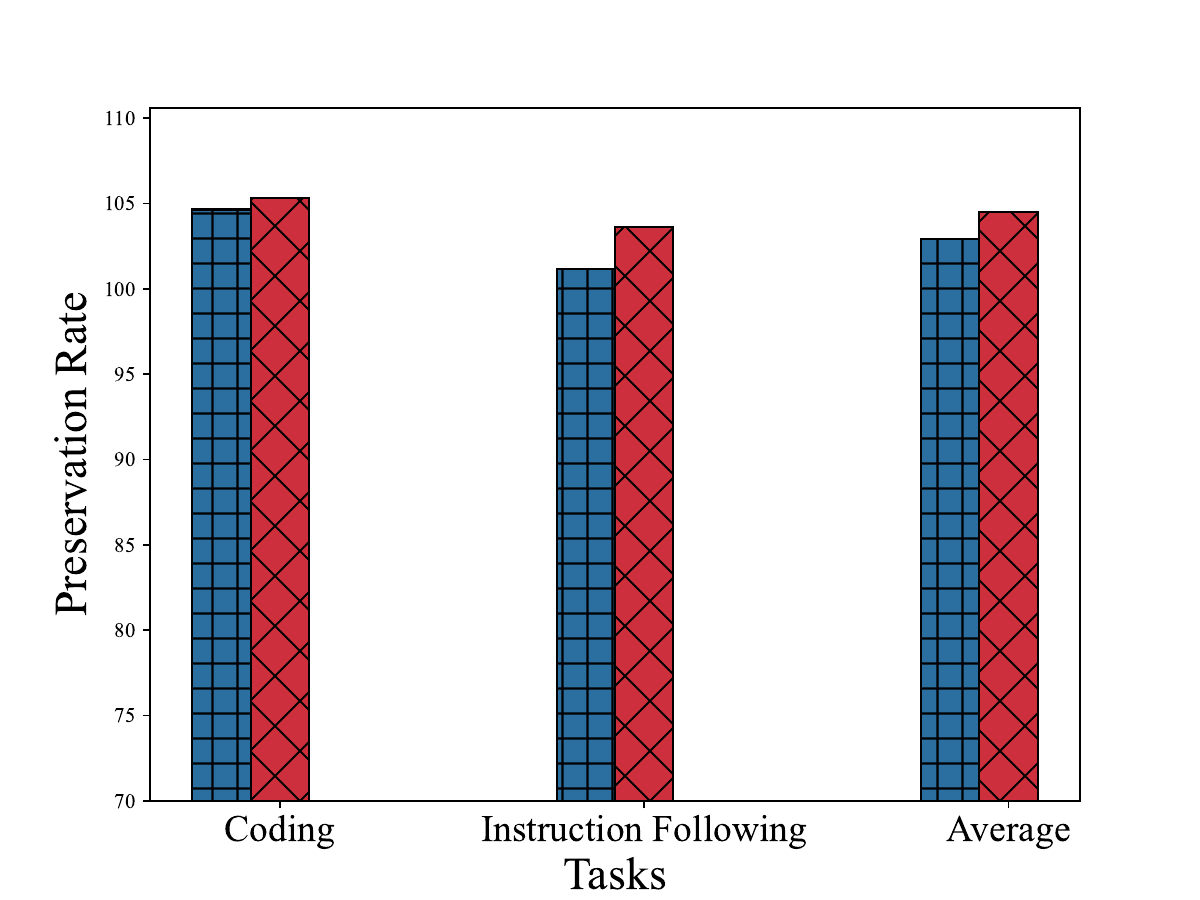}%
      \label{fig:addition_preservation_Llama}
  }
  \hspace{0.02\textwidth}
  \subfigure[CCT]{%
      \includegraphics[width=0.3\textwidth]{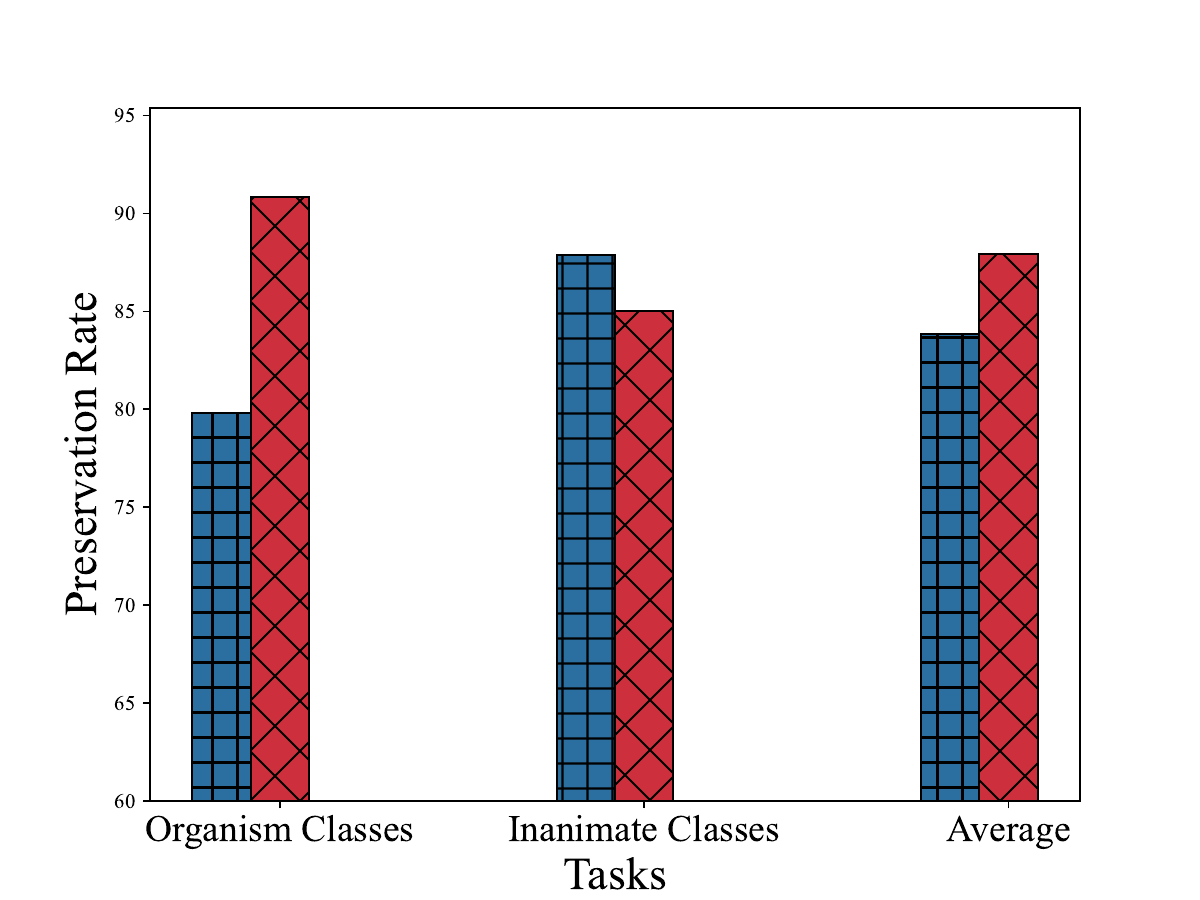}%
      \label{fig:addition_preservation_CCT}
  }
  \hspace{0.02\textwidth}
  \subfigure[Interfuser]{%
      \includegraphics[width=0.3\textwidth]{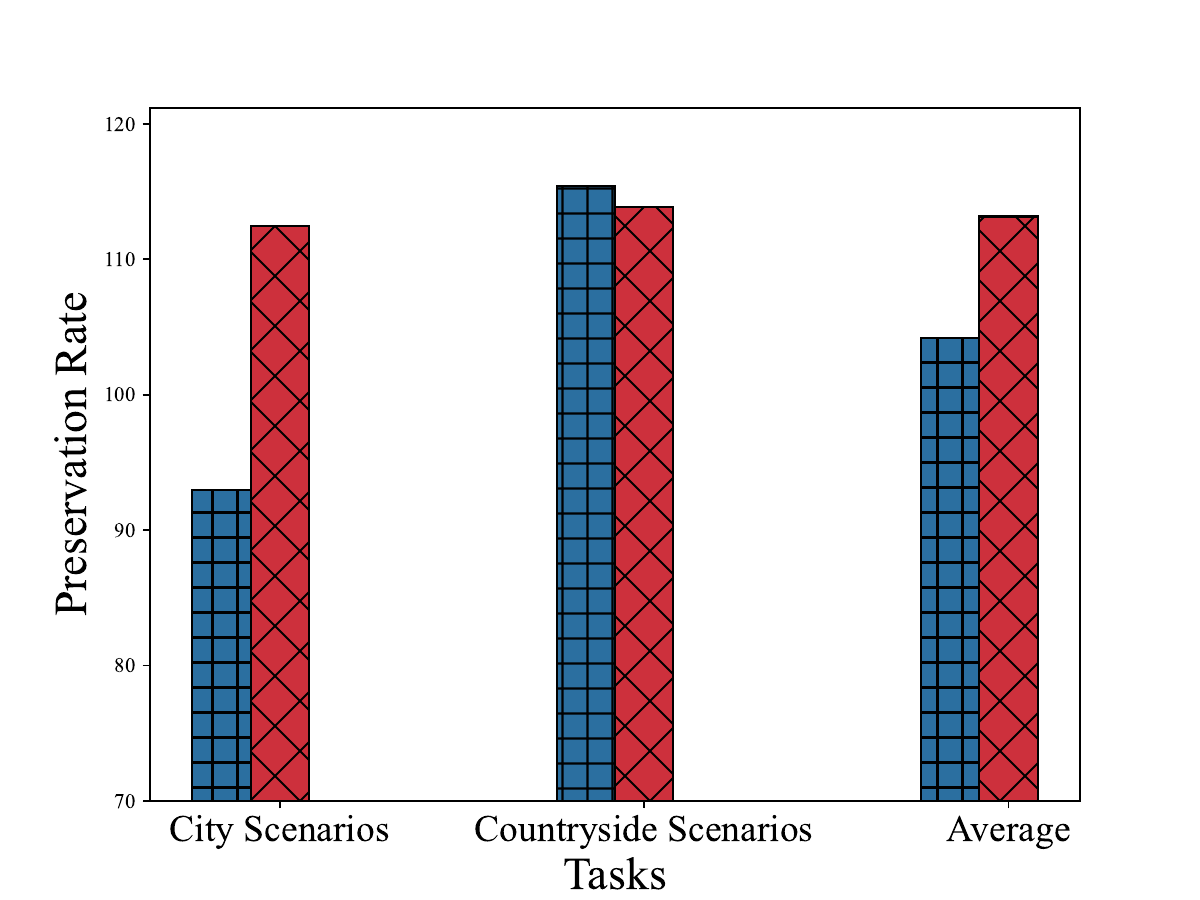}%
      \label{fig:addition_preservation_Interfuser}
  }
  \vspace{-5pt}
  \caption{Preservation Rate Comparision of \tool with Manual Grid Search}
  \label{fig:addition_comparision_effectiveness}
\end{figure}
Besides, Table~\ref{tab:ads_baymerge_results} shows the effectiveness of \tool on the \textit{Interfuser} architecture in autonomous driving domain. 
Specifically, $Interfuser_{\tool}$ achieves a $PR$ of 116.66\% in city scenarios, and a $PR$ of 113.77\% in countryside scenarios, with a $PD$ of 2.89\%. $Interfuser_{\tool}$ surpasses the source models on all the metrics, except for the \textit{Infraction Penalty} metric in countryside scenarios compared to \textit{Interfuser-Countryside}.

Fig. \ref{fig:comparision_effectiveness} shows the capability preservation rate of \tool compared to other model merging techniques when employing to the \textit{Llama2}, \textit{CCT}, and \textit{Interfuser} architectures. 
For the \textit{Llama2} architecture, \tool improves the $PR$ on coding and instruction following tasks, on average, by 16.70\% and 10.67\% respectively, and the average $PR$ across the two tasks by 13.46\%, compared to other model merging techniques.
For the \textit{CCT} architecture, \tool enhances the $PR$~on organism and inanimate class tasks by an average of 30.72\% and 13.47\%, respectively, and improves the overall average $PR$ across both tasks by 22.03\%, compared with other model merging techniques.
Compared with the best model, \ie $CCT_{\textsc{Linear}}$, among the models merged by the other five techniques, $CCT_{\textsc{\tool}}$ still achieves an improvement of $PR$ by 14.39\% and 3.06\% on organism and inanimate class tasks, respectively.
For the \textit{Interfuser} architecture, \tool outperforms all the model merging techniques on both city and countryside scenarios, with an average $PR$ improvement of 63.56\% across the two tasks. In addition, \tool achieves an average $PD$ decrease of 51.94\% compared to other model merging techniques across the three model architectures and~domains.

\begin{tcolorbox}[size=small, opacityfill=0.15, before skip=10pt, after skip=10pt]
  \textit{\textbf{Summary.}} \tool can effectively obtain an outperforming multi-task merged model~from task-specific source models across different domains, achieving higher capability preservation rate and lower preservation discrepancy compared to previous model merging techniques.
\end{tcolorbox}


\subsection{Efficiency Evaluation (RQ4)}

According to statistics, merging \textit{Llama2-7B-Code} and \textit{Llama2-7B-Chat} with \tool requires 31.18 hours, with a peak GPU memory consumption of 19.04 GB and a GPU utilization of 85\%,~consuming the most resources among all our model merging processes. This is mainly because the \textit{Llama2} architecture has the largest number of parameters, and the inference of large language models is time-consuming, which prolongs the evaluation phase during the search process. Merging \textit{CCT-Organism} and \textit{CCT-Inanimate} models takes 6.07 hours, reaching a peak GPU memory consumption of 4.68 GB and a GPU utilization of 82\%. When merging \textit{Interfuser-City} and \textit{Interfuser-Countryside}, \tool required 6.34 hours, with a peak GPU memory consumption of 1.46 GB and a GPU utilization of 75\%.

{To demonstrate the efficiency of \tool, we compare it with manual grid search, which evaluates all (\textit{weight}, \textit{density}) combinations as performed in RQ2. For the \textit{CCT} architecture, manual grid search requires 24,840 seconds (approximately 6.9 hours) to evaluate all 36 combinations (6 \textit{weight} values $\times$ 6 \textit{density} values), while \tool completes the search in 6.07 hours, achieving a 13.6\% time reduction. For the \textit{Interfuser} architecture, manual grid search requires 69,930 seconds (approximately 19.43 hours), which is 3.06$\times$ longer than \tool's 6.34 hours. Although the computational resource usage (GPU memory and utilization) is similar between manual grid search and \tool, \tool achieves significant time savings by intelligently selecting evaluation points through Bayesian optimization rather than exhaustively evaluating all combinations. This efficiency gain becomes more pronounced as the search space grows, making \tool a practical solution for model merging in resource-constrained environments.}

\begin{tcolorbox}[size=small, opacityfill=0.15, before skip=10pt, after skip=10pt]
    \textit{\textbf{Summary.}} \tool achieves significant time savings compared to manual grid search, demonstrating its efficiency in model merging.
\end{tcolorbox}


\subsection{Ablation Study (RQ5)}

\begin{table}[!t]
    \caption{Ablation Results of Effectiveness on \textit{CCT} Architecture in Image Classification}
    \vspace{-5pt}
    \centering
    \label{tab:cct_ablation_results}
    \begin{adjustbox}{width=0.5\textwidth}
    \begin{tabular}{cccccccccccccccc}
      \toprule
      \multirow{2}{*}{Model}& \multicolumn{2}{c}{Organism Classes} & \multicolumn{2}{c}{Inanimate Classes}\\
      \cmidrule(lr){2-3} \cmidrule(lr){4-5}
       & Top@1 $\uparrow$ & Top@5 $\uparrow$ & Top@1 $\uparrow$ & Top@5 $\uparrow$ \\
      \midrule
      \(CCT_{\textsc{\tool}}\) & \textbf{47.36} & \textbf{70.45} & \textbf{34.70} & \textbf{56.76} \\
      \textit{w/o segmentor} & 27.95 & 56.08 & 27.30 & 52.36 \\
      \bottomrule
    \end{tabular}    
    \end{adjustbox}
\end{table}

\begin{table}[!t]
  \caption{Ablation Results of Effectiveness on \textit{Interfuser} Architecture in Autonomous Driving}
  \vspace{-5pt}
  \centering
  \label{tab:ads_ablation_results}
  \begin{adjustbox}{width=\textwidth}
  \begin{tabular}{c*{6}{c}}
    \toprule
    \multirow{2}{*}{Model} & 
    \multicolumn{3}{c}{City Scenarios} & 
    \multicolumn{3}{c}{Countryside Scenarios} \\
    \cmidrule(lr){2-4} \cmidrule(lr){5-7}
    & Route Completion $\uparrow$ & Infraction Penalty $\uparrow$ & Driving Score $\uparrow$ & 
     Route Completion $\uparrow$ & Infraction Penalty $\uparrow$ & Driving Score $\uparrow$ \\
    \midrule
    \(Interfuser_\text{\tool}\) & \textbf{96.00} & \textbf{1.09} & \textbf{89.51} & 59.99 & \textbf{0.65} & \textbf{47.83} \\
    \textit{w/o segmentor} & 71.78 & 0.94 & 70.81 & \textbf{60.33} & 0.56 & 42.07 \\
    \bottomrule
  \end{tabular}    
  \end{adjustbox} 
\end{table}

\begin{figure}[!t]
  \centering
  
  \subfigure[GPU Memory Usage]{%
      \includegraphics[width=0.3\textwidth]{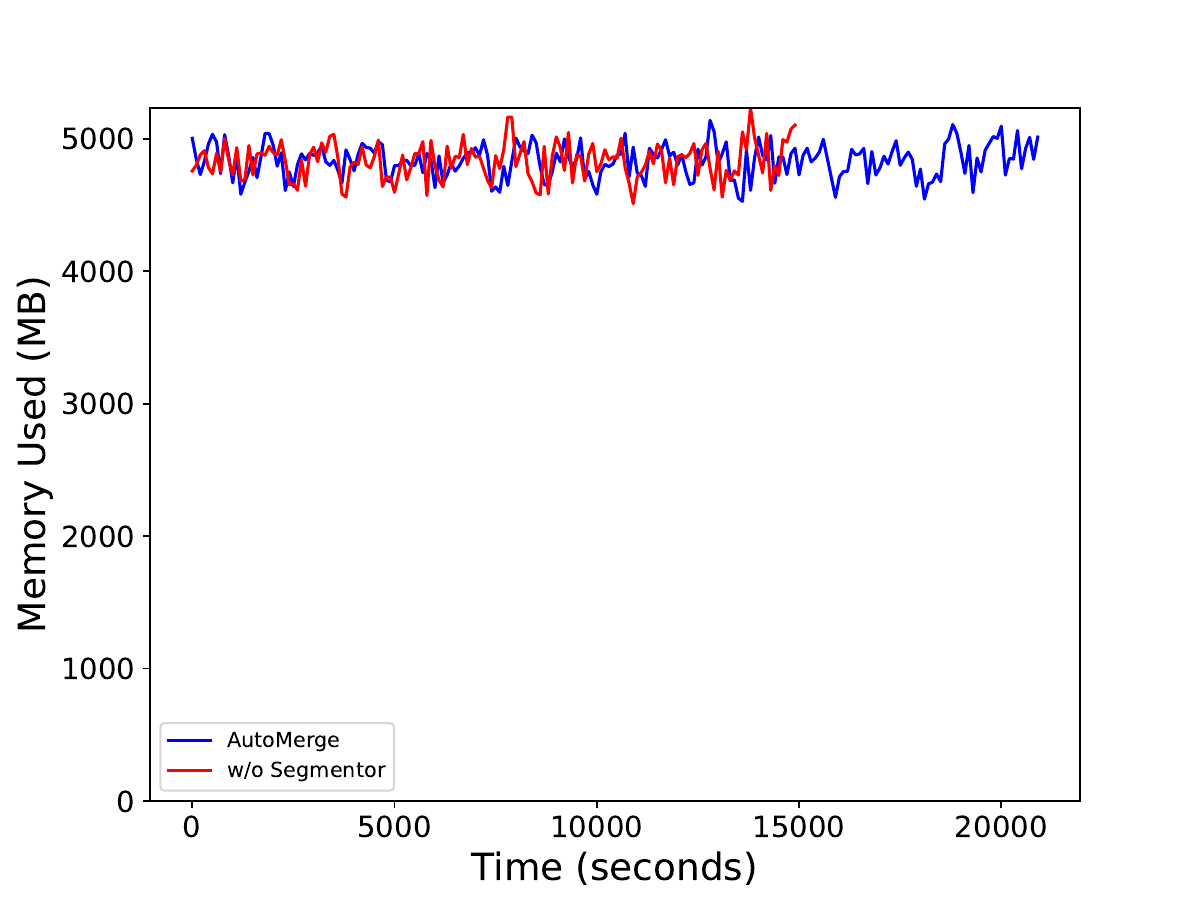}%
      \label{fig:cct_memory_usage}
  }
  \hspace{0.05\textwidth}
  \subfigure[GPU Utilization]{%
      \includegraphics[width=0.3\textwidth]{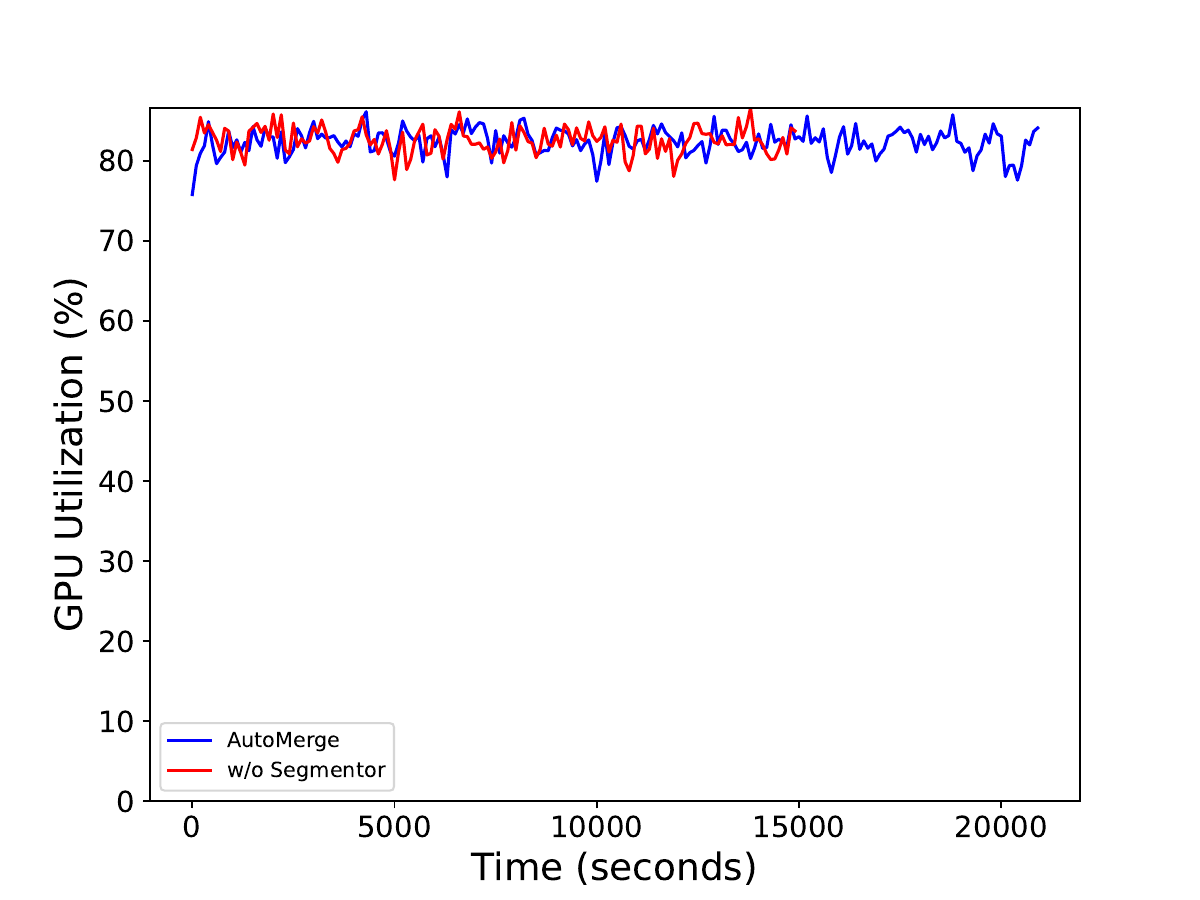}%
      \label{fig:cct_utilization}
  }

  \vspace{-5pt}
  \caption{Ablation Results of Efficiency on \textit{CCT} Architecture in Image Classification}
  \label{fig:cct_efficiency}
\end{figure}

\begin{figure}[!t]
  \centering
  
  \subfigure[GPU Memory Usage]{%
      \includegraphics[width=0.3\textwidth]{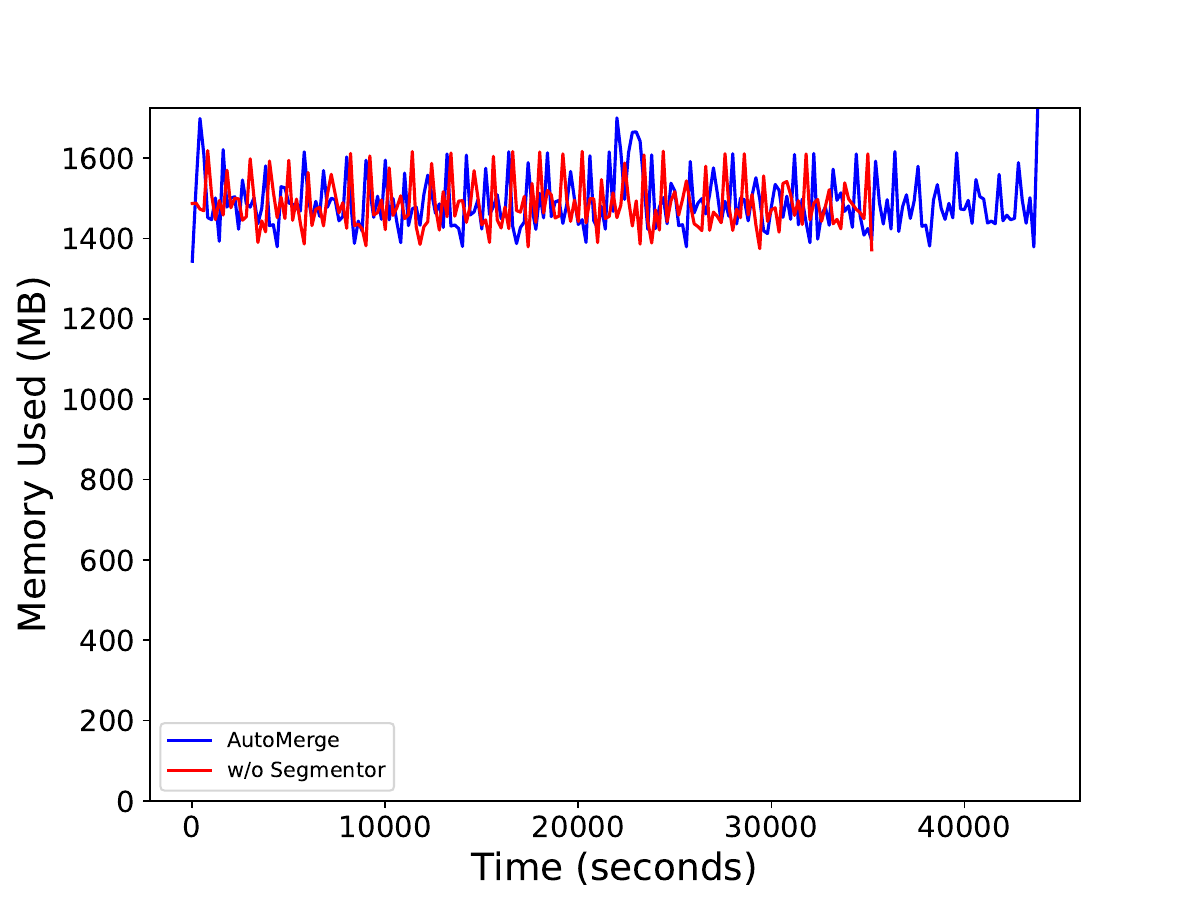}%
      \label{fig:interfuser_memory_usage}
  }
  \hspace{0.05\textwidth}
  \subfigure[GPU Utilization]{%
      \includegraphics[width=0.3\textwidth]{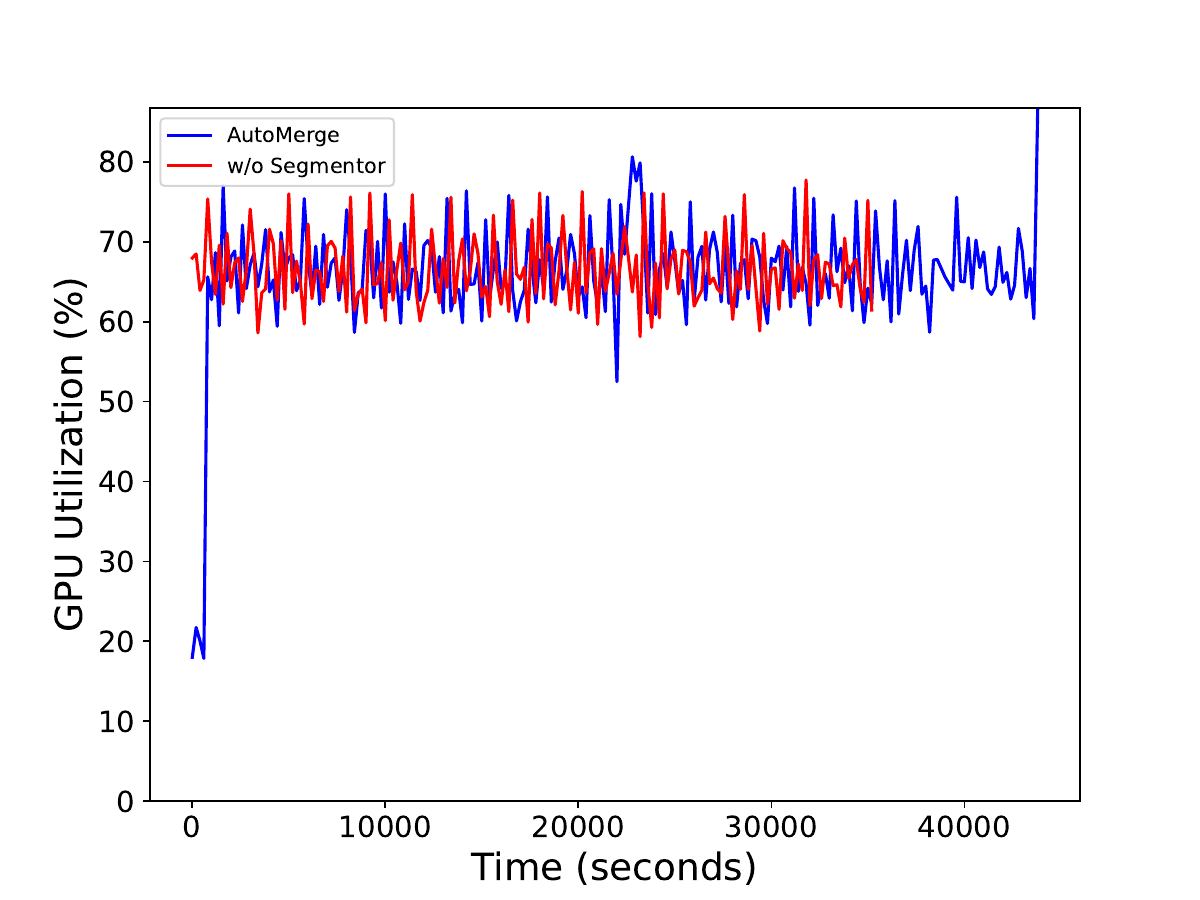}%
      \label{fig:interfuser_utilization}
  }

  \vspace{-5pt}
  \caption{Ablation Results of Efficiency on \textit{Interfuser} Architecture in ADS}
  \label{fig:interfuser_efficiency}
\end{figure}

\tool considers source models of the \textit{Llama2} architecture as a pair of entire Transformer blocks after segmentation, performing the search for the optimal merging configuration at a whole-model level. Consequently, removing the model segmentor does not affect the merging results of the \textit{Llama2} architecture. Therefore, we report ablation results on the \textit{CCT} architecture in image classification, and the \textit{Interfuser} architecture in autonomous driving domain.

Table~\ref{tab:cct_ablation_results} presents the ablation results on the \textit{CCT} architecture. The \textit{w/o segmentor} variant achieves a $PR$ of 62.84\% on organism classes and a $PR$ of 72.75\% on inanimate classes. In contrast, \tool demonstrates clear improvements. It raises the $PR$ on the organism classes task by 44.55\%, and on the inanimate classes task by 16.99\%. 
Similarly, Table~\ref{tab:ads_ablation_results} reports the ablation results on the~\textit{Interfuser} architecture. Under the \textit{w/o segmentor} setting, the merged model achieves a $PR$ of 93.76\% in city scenarios and 86.95\% in countryside scenarios. When compared with this variant, \tool delivers further gains, improving the $PR$ by 19.92\% in city scenarios and by 30.97\% in countryside scenarios. These results collectively demonstrate that segmentation plays a crucial role in enhancing the effectiveness of the capability preservation of source models in the merged models.

Fig.~\ref{fig:cct_efficiency} presents the ablation results of efficiency on the \textit{CCT} architecture in terms of peak~GPU memory usage and the GPU utilization. \tool has a higher time consumption than the \textit{w/o segmentor} variant, taking approximately 1.91 more hours, on average, to complete the merging process. However, the GPU memory usage, and the GPU utilization fluctuations are similar between \tool and the \textit{w/o segmentor} variant, suggesting that the computational load on the GPU remains largely unchanged. 
Fig.~\ref{fig:interfuser_efficiency} reports the ablation results of efficiency on the \textit{Interfuser} architecture, which are similar to those on the \textit{CCT} architecture.

\begin{tcolorbox}[size=small, opacityfill=0.15, before skip=10pt, after skip=10pt]
  \textit{\textbf{Summary.}} Searching for an optimal merging configuration at the block-level by incorporating segmentation into the merging process substantially 
  enhances the preservation of source model capabilities across both the \textit{CCT} and \textit{Interfuser} architectures, with acceptable time overhead compared to the whole-model merging.
\end{tcolorbox}


\subsection{Practical Applicability (RQ6)}

\begin{table}[!t]
    \caption{Effectiveness Comparison between \tool and Full Fine-tuning on \textit{Llama2} Architecture}
    \vspace{-5pt}
    \centering
    \label{tab:llm_retraining_results}
    \begin{adjustbox}{width=\textwidth}
    \begin{tabular}{cccccccccccccccc}
      \toprule
      \multirow{2}{*}[-11pt]{\specialcell{Model}} & 
      \multicolumn{6}{c}{\textbf{HumanEvalPack}} & 
      \multicolumn{5}{c}{\textbf{MMLU-Pro}} \\ 
      \cmidrule(lr){2-7} \cmidrule(lr){8-12} 
      & \multicolumn{3}{c}{Code Explanation (Pass@10)} & 
      \multicolumn{3}{c}{Code Synthesis (Pass@10)} & 
      \multicolumn{5}{c}{Instruction Following (Accuracy)} \\ 
      \cmidrule(lr){2-4} \cmidrule(lr){5-7} \cmidrule(lr){8-12} 
      & {\rotatebox[origin=c]{0}{Python $\uparrow$}} 
      & {\rotatebox[origin=c]{0}{Java $\uparrow$}} 
      & {\rotatebox[origin=c]{0}{JavaScript $\uparrow$}} 
      & {\rotatebox[origin=c]{0}{Python $\uparrow$}} 
      & {\rotatebox[origin=c]{0}{Java $\uparrow$}} 
      & {\rotatebox[origin=c]{0}{JavaScript $\uparrow$}} 
      & {\rotatebox[origin=c]{0}{Biology $\uparrow$}} 
      & {\rotatebox[origin=c]{0}{Business $\uparrow$}} 
      & {\rotatebox[origin=c]{0}{Chemistry $\uparrow$}} 
      & {\rotatebox[origin=c]{0}{Science $\uparrow$}} 
      & {\rotatebox[origin=c]{0}{Economics $\uparrow$}} \\ 
      \midrule
      \( Llama2_\text{\textsc{\tool}} \)  & \textbf{22.43} & \textbf{26.34} & \textbf{18.29} & \textbf{45.12} & \textbf{48.78} & \textbf{42.03} & 42.26 & \textbf{23.31} & \textbf{17.67} & 23.17 & \textbf{35.31} \\
      \( Llama2_\text{\textsc{retraining}} \)  & 20.31 & 23.58 & \textbf{18.29} & 41.77 & 48.36 & 38.74 & \textbf{42.29} & \textbf{23.31} & \textbf{17.67} & \textbf{23.29} & \textbf{35.31} \\
      \bottomrule
    \end{tabular}    
\end{adjustbox} 
\end{table}

\begin{table}[!t]
    \caption{Effectiveness Comparison between \tool and Full Fine-tuning on \textit{CCT} Architecture}
    \vspace{-5pt}
    \centering
    \label{tab:cct_retraining_results}
    \begin{adjustbox}{width=0.5\textwidth}
    \begin{tabular}{cccccccccccccccc}
      \toprule
      \multirow{2}{*}{Model}& \multicolumn{2}{c}{Organism Classes} & \multicolumn{2}{c}{Inanimate Classes}\\
      \cmidrule(lr){2-3} \cmidrule(lr){4-5}
       & Top@1 $\uparrow$ & Top@5 $\uparrow$ & Top@1 $\uparrow$ & Top@5 $\uparrow$ \\
      \midrule
      \(CCT_\text{\tool}\) & 47.36 & 70.45 & 29.80 & 52.76 \\
      \(CCT_{\textsc{training}}\) & \textbf{47.88} & \textbf{71.37} & \textbf{30.74} & \textbf{60.44} \\
      \bottomrule
    \end{tabular}    
    \end{adjustbox}
\end{table}

\begin{table}[!t]
    \caption{Effectiveness Comparison between \tool and Full Fine-tuning on \textit{Interfuser} Architecture}
    \vspace{-5pt}
    \centering
    \label{tab:ads_retraining_results}
    \begin{adjustbox}{width=\textwidth}
    \begin{tabular}{c*{6}{c}}
      \toprule
      \multirow{2}{*}{Model} & 
      \multicolumn{3}{c}{City Scenarios} & 
      \multicolumn{3}{c}{Countryside Scenarios} \\
      \cmidrule(lr){2-4} \cmidrule(lr){5-7}
      & Route Completion $\uparrow$ & Infraction Penalty $\uparrow$ & Driving Score $\uparrow$ & 
       Route Completion $\uparrow$ & Infraction Penalty $\uparrow$ & Driving Score $\uparrow$ \\
      \midrule
      \(Interfuser_\text{\tool}\) & \textbf{96.00} & \textbf{1.09} & \textbf{89.51} & 59.99 & 0.65 & 47.83 \\
      \(Interfuser_{\textsc{training}}\) & 95.34 & 1.06 & 89.25 & \textbf{60.13} & \textbf{0.67} & \textbf{48.78} \\
      \bottomrule
    \end{tabular}    
\end{adjustbox} 
\end{table}

For the \textit{Llama2} architecture, as shown in Table~\ref{tab:llm_retraining_results}, the model merged by \tool achieves~capabilities that are comparable to the fully fine-tuned model across both coding and instruction-following tasks. While the fully fine-tuned \textit{Llama2} model attains slightly higher scores in most metrics, \tool yields only marginal drops, suggesting that \tool can preserve the majority of task-specific knowledge without retraining in LLMs.

For the \textit{CCT} architecture, as shown in Table~\ref{tab:cct_retraining_results}, \tool achieves nearly identical accuracy on organism classes and only slightly lower capability on inanimate classes with only marginal drops of 4.50\%, compared to the fully fine-tuned model. This indicates that \tool can preserve the majority of task-specific knowledge without retraining in image classification.

For the \textit{Interfuser} architecture, as shown in Table~\ref{tab:ads_retraining_results},
$\textit{Interfuser}_\tool$ delivers nearly identical capabilities to the fully fine-tuned model in both city and countryside driving scenarios. Differences in \textit{Route Completion} and \textit{Driving Score} are within 2 percentages, while \textit{Infraction Penalty} remains almost unchanged, underscoring the effectiveness of \tool in autonomous driving.

\begin{table}[!t]
  \vspace{-5pt}
  \centering
  \caption{Efficiency Comparison between \tool and Full Fine-tuning on Different Model Architectures}
  \label{tab:efficiency_comparison_reuse_approaches}
  \begin{adjustbox}{width=0.7\textwidth}
    \begin{tabular}{ccccccccccc}
    \toprule
    \multirow{2}{*}{Model} & \multicolumn{3}{c}{Consumption Time ($h$) $\downarrow$} & \multicolumn{3}{c}{GPU Memory Usage ($GB$) $\downarrow$} & \multicolumn{3}{c}{GPU Utilization (\%) $\uparrow$} \\
    \cmidrule(lr){2-4} \cmidrule(lr){5-7} \cmidrule(lr){8-10}
    & \textit{Llama2} & \textit{CCT} & \textit{Interfuser} & \textit{Llama2} & \textit{CCT} & \textit{Interfuser} & \textit{Llama2} & \textit{CCT} & \textit{Interfuser} \\
    \midrule
    \textsc{\tool} & \textbf{31.18} & \textbf{6.07} & \textbf{6.34} & \textbf{19.04} & \textbf{4.68} & \textbf{1.46} & \textbf{85} & 82 & 75 \\
    \textsc{Full Fine-tuning} & 82.63 & 40.15 & 10.60 & 23.42 & 22.04 & 21.85 & \textbf{85} & \textbf{85} & \textbf{80} \\
    \bottomrule
    \end{tabular}
  \end{adjustbox}
  \end{table}

Table~\ref{tab:efficiency_comparison_reuse_approaches} presents the efficiency comparison between \tool and full fine-tuning on different model architectures. Across all three model architectures, \tool requires significantly less training time. \tool saves the time consumption for obtaining a multi-task model by 62.43\%, on average, compared to full fine-tuning. Besides, the benefits extend to GPU memory consumption. While full fine-tuning incurs high memory overheads, \tool reduces these requirements of computational resources drastically by 64.34\%, on average. Finally, \tool maintains efficiency levels comparable to full fine-tuning regarding GPU utilization.


\begin{tcolorbox}[size=small, opacityfill=0.15, before skip=10pt, after skip=10pt]
  \textit{\textbf{Summary.}} Compared to full fine-tuning, \tool achieves multi-task models with capabilities closely aligned with retraining-based counterparts across different domains, and avoids the substantial cost of retraining, promoting the model reuse of task-specific models.
\end{tcolorbox}


\subsection{Threats to Validity}

First, the selection of model architectures and domains poses a threat to validity. To this end,~we select three distinct model architectures, \ie \textit{Llama2}, \textit{CCT}, and \textit{Interfuser}, across three widely concerned domains, \ie LLMs, image classification, and autonomous driving. These model architectures are complex and representative, covering diverse inner blocks such as Transformer, CNN, MLP, and GRU. \tool shows great effectiveness and efficiency across these model architectures, and we believe \tool can be applied to other model~architectures across different domains.

Second, the selection of model merging techniques poses another threat to validity. To mitigate this threat, we select five state-of-the-art model merging techniques, covering both weight-based and subspace-based techniques. We do not choose other techniques because they were either not peer-reviewed or not open-source. All these techniques are orthogonal to \tool and can be integrated into our framework to obtain merged models with better performance.

Third, the randomness during the optimization process affects the validity. To mitigate this threat, we set the number of search iterations to 200 and repeat the experiment five times, reporting the average results. We have conducted a paired t-test analysis~\cite{box1987guinness} across the metrics of the merged models obtained by \tool and the five model merging techniques, and \tool achieves average p-values of 0.026, indicating statistically significant improvements~(p$<$0.05).

Last, \tool currently supports merging only task-specific source models with identical architectures to form a multi-task model. This constraint arises from the inherent challenges in merging models with different architectures, which is a common shortcoming shared by existing model merging techniques. However, we believe that as model merging techniques continue to evolve, we can integrate more advanced techniques into our framework, enabling the merging of source models with different architectures. This advancement would further drive forward the reuse of diverse model architectures across multiple tasks.


\section{SE Discussion}


\section{Related Work}
\textbf{Model Merging.}
Considering both the difficulty of obtaining training data and the cost of retraining for knowledge transfer, model merging~\cite{li2023deep,yang2024model} has been widely explored to integrate multiple pre-trained task-specific models with the same architecture into a single model without additional training within the LLMs domain. These techniques can be broadly categorized into weight-based techniques~\cite{utans1996weight,ilharco2022editing,matena2022merging,yang2023adamerging}, which leverage various rules to formally merge the parameter weights, and subspace-based techniques~\cite{yu2024language,yadav2023ties,yang2024model,huang2024emr}, which rebuild the models into sparse subspace for reducing conflict during merging. Specifically, Utans et al.~\cite{utans1996weight} propose the first model merging technique, \ie \textsc{Linear}, by simply averaging models' parameter weights. The \textsc{Task Arithmetic}~\cite{ilharco2022editing} technique introduces the concept of task vector to measure the difference between the fine-tuned model and the base model, further improving the performance of the merged model. Based on the task vector, Yadav et al.~\cite{yadav2023ties} add a coefficient to prune redundant parameters to solve the conflict of signs during the merging process. Recently, the \textsc{DARE}~\cite{yu2024language} technique has been proposed to randomly set some delta parameters to zero, and then rescale the remaining parameters, improving the effectiveness of merging. AKiba et al.~\cite{akiba2025evolutionary} use an evolutionary algorithm to optimize the selection of merging coefficients, optimizing the model inference path to improve model performance. Our work aims to investigate the generalization of these techniques to complex model architectures in different domains, and these techniques are orthogonal to \tool and can be integrated into our framework to obtain merged models with better performance. Su et al.~\cite{su2025fine} propose the closest work to \tool. They randomly split the Transformer layers in LLMs and adopt different merging configurations for different layers at a more fine-grained level. However, this approach is only applicable to models that fully use the Transformer architecture.
In contrast, \tool searches the merging configurations at the block level, which can be applied to more complex model architectures in different domains, promoting broader~model~reuse. 

\textbf{Model Reuse.}
With the widespread adoption of task-specific pre-trained models, model reuse has emerged as a prominent research topic, extending the classical notion of software reuse. Various approaches have been proposed to facilitate model reuse when constructing a unified multi-task model, \eg model fine-tuning~\cite{hu2022lora,devlin2019bert,dodge2020fine,xu2023improving,tu2025robust}, MoE~\cite{shen2024efficient,tang2024merging}, and knowledge distillation~\cite{yang2025hyperbolic,agand2024knowledge,zhou2025all,gao2024complementary,cui2025multi,xiang2025dkdm,liu2024small}. Model fine-tuning is the most common approach for model reuse. The core idea behind this approach is to adapt a pre-trained model to tasks by updating its parameters using the training datasets of the tasks. Hu et al. propose the widely adopted LoRA~\cite{hu2022lora}, which reduces the number of parameters required for fine-tuning by introducing low-rank adapters. However, it may lead to unexpected training time consumption in the case of large-scale datasets~\cite{yao2025pre}. Tang et al.~\cite{tang2024merging} explore the effectiveness of MoE in multi-task learning. But, MoE depends on complex expert selection mechanisms and typically enlarges the overall model~\cite{taraghi2024deep}. Knowledge distillation is an approach that transfers knowledge from a teacher model to a student model. Thadajarassiri et al.~\cite{thadajarassiri2023knowledge} propose a multi-task learning framework based on distillation of pre-trained models, enhancing the multi-task capability of the distilled model. However, Cui et al.~\cite{cui2025multi} point out that when task-specific data is limited, information loss during the distillation process may lead to unstable performance in the student model. Overall, these approaches remain constrained by  capability decay~\cite{waheed2024distill,puigcerver2022adversarial, zhang2023robust} and high training time consumption~\cite{yao2025pre,taraghi2024deep}. In contrast, \tool adopts a search-based framework that applies model merging techniques across diverse architectures and domains, facilitating model reuse of multiple task-specific models.


\section{Conclusion}
We conduct the first empirical study of model merging across diverse model architectures and domains, revealing critical limitations that hinder the effectiveness of current techniques. Inspired by the insights, we propose a search-based framework, \tool, to support model merging techniques on different model architectures and domains, ensuring the performance of merged models while reducing time consumption, offering an effective and efficient perspective on model~reuse.

\section{Data Availability} All the experimental data and source code of our work is available at our replication site~\cite{website}.


\bibliographystyle{ACM-Reference-Format}
\bibliography{src/reference}

\end{document}